\documentclass{article}
\usepackage{natbib,latexsym,epsfig,emulateapj,onecolfloat,graphics}
\def\eps@scaling{.95}
\def\epsscale#1{\gdef\eps@scaling{#1}}

\def\plotone#1{\centering \leavevmode
    \epsfxsize=\eps@scaling\columnwidth \epsfbox{#1}}

\def\kms{\ifmmode {\rm\,km\,s^{-1}}\else
    ${\rm\,km\,s^{-1}}$\fi}
\def\ms{\ifmmode {\rm\,m\,s^{-1}}\else
    ${\rm\,m\,s^{-1}}$\fi}
\def\kmsMpc{\ifmmode {\rm\,km\,s^{-1}\,Mpc^{-1}}\else
    ${\rm\,km\,s^{-1}\,Mpc^{-1}}$\fi}
\def\hkmsMpc{\ifmmode {\rm\,h^{-1}\,km\,s^{-1}\,Mpc^{-1}}\else
    ${\rm\,h^{-1}\,km\,s^{-1}\,Mpc^{-1}}$\fi}
\def\lya{{\rm Ly}$\alpha$}

\def\kpc{{\rm\,kpc}}

\def\msun{\ifmmode {\rm\,M_\odot}\else ${\rm\,M_\odot}$\fi}
\def\Msun{\ifmmode {\rm\,M_\odot}\else ${\rm\,M_\odot}$\fi}
\def\lsun{\ifmmode {\rm\,L_\odot}\else ${\rm\,L_\odot}$\fi}
\def\Lsun{\ifmmode {\rm\,L_\odot}\else ${\rm\,L_\odot}$\fi}
\def\rsun{\ifmmode {\rm\,R_\odot}\else ${\rm\,R_\odot}$\fi}
\def\Rsun{\ifmmode {\rm\,R_\odot}\else ${\rm\,R_\odot}$\fi}
\def\pc{{\rm\,pc}}

\def\cmtw{\ifmmode {\rm\,cm^{-2}}\else ${\rm\,cm^{-2}}$\fi}
\def\cmthr{\ifmmode {\rm\,cm^{-3}}\else ${\rm\,cm^{-3}}$\fi}
\def\yr{{\rm\,yr}}

\def\ergps{\ifmmode {\rm\,erg\,s^{-1}}\else ${\rm\,erg\,s^{-1}}$\fi}
\def\ergpscmtw{\ifmmode {\rm\,erg\,cm^{-2}\,s^{-1}}}
\def\Junits{\ifmmode {\rm\,erg\,cm^{-2}\,s^{-1}\,Hz^{-1}\,sr^{-1}}}

\def\cf{{\it cf.~}\ }
\def\eg{{\it e.g.}}

\def\deg{\ifmmode {^{\circ}}\else {$^\circ$}\fi}
\def\degr{\ifmmode {^{\circ}}\else {$^\circ$}\fi}
\def\degs{\ifmmode {^{\circ}}\else {$^\circ$}\fi}

\def\Ho{\ifmmode {\rm\,H_\circ}\else ${\rm\,H_\circ}$\fi}
\def\hnot{\ifmmode {\rm\,H_\circ}\else ${\rm\,H_\circ}$\fi}
\def\h0{\ifmmode {\rm\,H_\circ}\else ${\rm\,H_\circ}$\fi}
\def\hnotunit{\ifmmode {\rm\,km\,s^{-1}\,Mpc^{-1}}\else
    ${\rm\,km\,s^{-1}\,Mpc^{-1}}$\fi}
\def\qnot{\ifmmode {\rm\,q_\circ}\else ${\rm q_\circ}$\fi}
\def\q0{\ifmmode {\rm\,q_\circ}\else ${\rm q_\circ}$\fi}
\def\ie{{\it i.e.}}

\def\vs{{\it versus} }

\def\arcsec{\ifmmode {^{\prime\prime}}\else $^{\prime\prime}$\fi}
\def\asec{\ifmmode {^{\prime\prime}}\else $^{\prime\prime}$\fi}
\def\arcmin{\ifmmode {^{\prime}}\else $^{\prime}$\fi}
\def\amin{\ifmmode {^{\prime}}\else $^{\prime}$\fi}

\def\heone{He {\small I}}
\def\hetwo{He {\small II}}
\def\hone{H {\small I}}
\def\h{{\rm h}}
\def\lesssim{\mathrel{\hbox{\rlap{\hbox{\lower4pt\hbox{$\sim$}}}\hbox{$<$}}}}
\def\gtrsim{\mathrel{\hbox{\rlap{\hbox{\lower4pt\hbox{$\sim$}}}\hbox{$>$}}}}
\let\la=\lesssim                        
\let\ga=\gtrsim

\def\be{\begin{equation}}
\def\ee{\end{equation}}

\lefthead{C. V. Manning}
\righthead{Modeling the Void \hone\ Column Density Spectrum }

\begin{document}
\twocolumn [ 

\title{Modeling the Void \hone\ Column Density Spectrum \\
with Sub-Galactic Halos}
\author{Curtis V. Manning}
\affil{Astronomy Department, University of California, Berkeley, CA
94720} 
\authoremail{cmanning@astro.berkeley.edu}

\begin{abstract}

The equivalent width distribution function (EWDF) of \hone\ absorbers
specific to the void environment has been recently derived
\citep{Manning:02}.  The findings revealed void line densities
$d{\mathcal N}/dz\simeq 500$ at equivalent widths ${\mathcal W} \ge
15.8$ m\AA\ ($N_{HI}\ga 2.6\times 10^{12} ~\cmtw$).  I show that the
void absorbers cannot be diffuse (or so-called filamentary) clouds,
expanding with the Hubble flow, as suggested by N-body/hydro
simulations.  Absorbers are here modeled as the baryonic remnants of
sub-galactic perturbations that have expanded away from their dark
halos in response to reionization at $z \approx 6.5$.  A 1-D
Lagrangian hydro/gravity code is used to follow the dynamic evolution
and ionization structure of the baryonic clouds for a range of halo
circular velocities.  The simulation products at $z=0$ can be combined
according to various models of the halo velocity distribution function
to form a column density spectrum that can be compared with the
observed \citep{Manning:02}.  To explain the observations with these
models requires a search of parameter space somewhat beyond the
envelope of convention.  For a given circular velocity, a halo model
more massive than the \citealt*{Navarro:96, Navarro:97} (NFW) halo is
required to reproduce the observed line density of absorbers.  A more
massive, non-singular isothermal halo is used with a more favorable
outcome.  I find that such clouds may explain the observed EWDF if the
halo velocity distribution function is as steep as that advanced by
\citet{Klypin:99}.  Observations are best explained when individual
halos have sub-halos that occupy the flanks of the ``parent''.  A
picture emerges in which growth by accretion of sub-halos is possible.
Further analysis suggests that the mass distribution about a cloud may
extend significantly farther than the virial radius.  Accounting for
the total void mass density remains an outstanding problem.

\end{abstract}


\newpage

 ]

\section{Introduction} \label{sec-intro}

Among the illuminati of cosmology, a paper that proposes to study
\lya\ clouds in voids, and further seeks to model them as remnants of
sub-galactic perturbations, is likely to be viewed as an anomaly;
N-body/hydro simulations have had such success in reproducing the
detailed nature of the \lya\ forest \citep[\eg,][]{MiraldaEscude:96,
Hernquist:96, Riediger:98, Cen:99, Dave:99, Dave:01}, that seriously
considering a different picture is nearly unthinkable.

In these simulations, \lya\ absorbers are \emph{not} physically
associated with discrete small halos; rather, they arise in large
diffuse sheets or filaments \citep[\eg,][]{Bi:93, Weinberg:97,
MiraldaEscude:96, Dave:99}.  This view is supported by the conclusions
of \citet{Dinshaw:97, Dinshaw:98} regarding the hit/miss statistics of
moderate redshift \lya\ absorbers in double or group quasars.  The
possibility that these hit/miss statistics could instead be the
signature of the clustering of small clouds was considered, and
rejected without a thorough analysis
\citep{Weinberg:97}\footnote{However, in unpublished work, Manning
(2000) produced a good fit to the hit/miss statistics over a range of
10 kpc to 1 Mpc using discrete \lya\ clouds clustered around small
galaxy groups of physical extent $\sim 750$ kpc (\eg, ``Local
Groups'').}.  Nor are \lya\ clouds expected to be found in voids at $z
\approx 0$ in any significant numbers, for in voids, the diffuse
filaments expand with the Hubble flow and disperse to below the
current limits of detectability at redshifts $z\la 1$
\citep{Riediger:98, Dave:99}.  Nevertheless, there exists now some
considerable history of detections of low-redshift \lya\ clouds in
voids \citep{Morris:93, Stocke:95, Shull:96, Tripp:98, Penton:00a,
McLin:02}, suggesting that perhaps void clouds are not as rare as the
simulations would imply.  Likewise, the discovery of compact high
velocity clouds (CHVCs) in the galactic neighborhood \citep{Braun:99},
which are successfully modeled as partially held by dark matter
\citep{Blitz:99, Sternberg:02}, suggests that sub-galactic halos with
associated \hone\ gas may not be as unlikely as previously thought.

At issue is not just the specific disposition of dark matter and
baryons in vast, dark, under-dense regions.  The observations of \lya\
clouds at low-redshift, where complete galaxy catalogs can be used to
calculate their relative isolation, may provide a crucial test of the
N-body simulations that play such a central role in informing our
picture of process on the cosmic scale.

Simulations have effectively been calibrated by the use of the
high-redshift \lya\ forest.  Under the assumption that dark matter
(DM) can be characterized by a 3-dimensional power spectrum $P(k)$
\citep{Bi:93, Croft:98}, and that gas traces DM except at the smallest
scales, where pressure forces are thought to be important
\citep{Bi:93}, $P(k)$ is derived from high-redshift \lya\ forest
spectra, and used to model the initial conditions of simulations.  The
\lya\ forest thus became a testing ground for simulations, enabling
researchers to gradually refine their code to a point that there is
now excellent agreement with observations of a wide range of \lya\
forest properties \citep{MiraldaEscude:96, Hernquist:96, Riediger:98,
Cen:99, Dave:99, Dave:01}.  These simulations show structure formation
and movement of matter from low, to high density regions in response
to peculiar gravitational fields.  For instance, \citet{Cen:99} use an
open CDM model to follow the transferral of matter from underdense, to
overdense regions, showing that over 90\% of the mass in overdense
(which we will later refer to as ``shocked'') regions at $z=0$ arrived
after $z=3$.

The challenging task of following the movement of gas in a simulated
cosmological context is handled by concentrating baryonic and dark
mass into baryon and dark matter ``particles'', typically of mass
$\sim 10^8$, and $\sim 10^9$ \Msun, respectively \citep{Katz:96,
Hernquist:96, Dave:99}.  Due to this coarseness, small galactic and
subgalactic halos cannot be represented.  Thus, the baryons that could
have been associated with smaller halos are distributed in a smooth
manner through the large diffuse structures.  However, the response of
the particles to gravity produces convergent motions which are
interpreted as absorption lines -- an effect known as the
``fluctuating Gunn-Peterson effect'' \citep{Croft:98}.  Artificial
spectra are typically extracted by smoothing groups of particles with
a spline kernel with a radial scale that is dependent on the density
\citep{Katz:96} so that gas resolution varies from $\sim 5$ to $\ga
200$ kpc for high, and low density regions, respectively
\citep{Hernquist:96}.  Thus, even if sub-galactic halos could be
resolved, the dynamics of the behavior of baryons in the halo could
not be realistically represented since the distribution of gas is
determined completely by the spline function, which has no physical
foundation.

In attempting to characterize the nature of the evolution of structure
seen in simulations, two broad types of environments are detected.  In
\citet{Riediger:98} these environments are characterized as
``shocked'' and ``unshocked'', while in \citet{Cen:99}, they are
``warm/hot'', and ``warm''.  In either case, the filling factors, and
the physics contributing to the physical state of the gas are
approximately the same.  The clouds in warm/hot, or shocked regions
are collapsing structures, and are shock heated, while the warm,
unshocked clouds characterize underdense regions in which gas is
expanding with the Hubble flow, and is hence adiabatically cooled.  In
\citet{Cen:99}, the collapsing regions are thought to comprise about
10\% of the volume of the universe at the present time.  In the
simulations of \citet{Dave:99} and \citet{Dave:01}, clouds at high redshift
populate both overdense and underdense regions, but for $z <1$,
absorbers are thought to arise in regions that are overdense by
factors of 12 to 100; stronger lines originating close to galaxies,
and weaker lines distributed in the diffuse filamentary structure.

At $z \sim 3$, the situation is different; the apparent lack of a
modulation in the incidence of \lya\ absorbers on velocity scales
appropriate to galaxy superclusters or voids in high-redshift spectra
prompted \citet{Carswell:87} to conclude that voids in the \lya\ cloud
distribution can fill no more than 5\% of the universe at $\langle z
\rangle \simeq 3.2$, a result corroborated by others.  However, the
area is murky (see the review by \citet{Rauch:98} for a discussion)
since how one defines a \lya\ cloud void depends on an arbitrary
detection threshold.

However, the movement of this proto-void material toward mass
concentrations will tend to evacuate them.  Simulations of voids by
\citet{Mathis:02} suggest that no galaxy of any size is expected to
currently populate voids of size $\sim 10~\h^{-1}$ Mpc.  However,
studies show that there are much larger voids, with an average
diameter greater than twice as large (\citealp[\eg,][]{Hoyle:02,
El-Ad:00}).  But void simulations of \cite{Arbabi:02} appear to imply
that larger voids have generally lower densities.  However, the
relatively large galaxy density within the huge Bo\"otes void
\citep{Dey:95}, and the subsequent detection of other galaxies in 21
cm emission \citep{Szomoru:96a, Szomoru:96b}, suggests that larger
voids may retain halos when smaller voids may not.

The assumption that baryons trace a diffuse DM distribution, and not
smaller, sub-galactic potentials, can, in principle be tested by the
search for \lya\ absorbers in local voids.  For, if absorbers are
diffuse, they will have expanded with the Hubble flow and now be so
diffuse that they are unobservable \citep[\eg][]{Reidiger:98}, while
if they are held by sub-galactic halos, the gravitational potentials
may be sufficient to restrain the evaporation of baryons and produce
absorbers.

Observations of low-redshift \lya\ clouds, aided by galaxy catalogs,
can be used to determine the relative isolation of clouds, and may
thus help to resolve these issues.  For instance, Hubble Space
Telescope (HST) spectra have detected numerous \hone\ clouds for which
no nearby galaxy was found \citep{Morris:91, Morris:93, Stocke:95,
Shull:96, Tripp:98, McLin:02}.  However, the methodology used by these
investigators, though useful for gaining insight into the low redshift
universe, does not facilitate the study of void clouds in any
systematic way.  For, ``nearest neighbor'' measures of a cloud's
isolation weights galaxies equally, whether dwarf or giant, and so
lacks the sensitivity to mass that is intuitively attached to the
concept of isolation.

A new method for resolving the environments of \lya\ clouds is
presented in \citet{Manning:02} (hereafter Paper 1).  It uses ambient
tidal fields, calculated from galaxy catalogs, as a relative measure
of a cloud's isolation, and thus weights the effects of galaxies by
their mass, and by the inverse cube of the distance.  Clouds can then
be sorted by their degrees of isolation.  A catalog of clouds can, for
instance, be divided into two; one sub-catalog with clouds abiding in
ambient tidal fields less than some arbitrary tide ${\mathcal
T}_{lim}$, and one with tide greater.  The analysis of sub-catalogs
into distribution functions can then be undertaken.  It was found that
clouds in low tidal field environments display significantly different
characteristics than those in high tidal field environments.

\begin{figure}[h!] 
\centering
\epsscale{1.1} 
\plotone{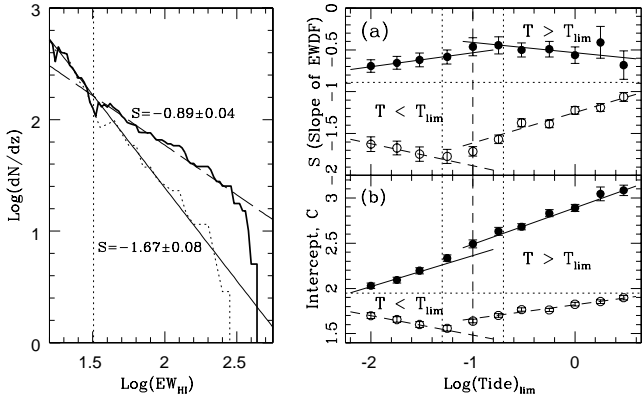}
\vspace{-3.6cm}
\caption{\label{fig-ewds2} The left-hand figure shows the log of the
mean cumulative line density (solid jagged line) as a function of the
log of the EW for low-$z$ Ly$\alpha$ clouds.  It is evident that there
is a broken power law, suggesting the existence of two populations of
clouds.  The spectral slopes, ${\mathcal S}$, are shown for weighted
linear fits to the mean EWDF for ${\mathcal W} \ge 32$ m\AA\ (dashed
line), and $\le 32$ m\AA\ (thin solid line).  The latter fit matches
the void EWDF slope well (dotted line).  The vertical dotted line is
at 32 m\AA.  The right-hand figure shows the trends of fitting
parameters, based on the model, $\log(d{\mathcal N}/dz) = C +
{\mathcal S} \log({\mathcal W}/63 {\rm m\AA})$, with tidal field
${\mathcal T}_{lim}$ for void, and non-void catalogs.  The upper panel
of this figure shows the slopes, and the lower, the intercepts (log of
line density at 63 m\AA).  The horizontal dotted lines show the slope
and intercept of the mean EWDF.  Upper and lower sub-panels show
non-void (tide as lower limit) and void (upper limit) EWDFs,
respectively.  The three vertical lines show the apparent range and
center of the ``transition zone'' (see text).  Note that
characteristic slopes of void EWDFs (right-hand plot) are in agreement
with the low EW slope of the mean EWDF (left-hand plot).}
\vspace{-0.2cm}
\end{figure}

When represented in a log-log plot, the equivalent width distribution
function (EWDF) is accurately fit by a power-law.  The left-hand
illustration of Fig.  \ref{fig-ewds2} shows the EWDF of \lya\ clouds,
using all data irrespective of tidal field (heavy solid line).  The
break in the distribution, seen at $\log({\mathcal W}) \simeq 1.5$
(${\mathcal W} \simeq 32$ m\AA, where ${\mathcal W}$ is the rest
equivalent width of the \lya\ absorber in m\AA) is suggestive of two
distinct populations of clouds represented in the data.  The thin
solid and dashed lines represent number weighted fits to the data at
${\mathcal W} \le 32$ m\AA, and ${\mathcal W} \ge 32$ m\AA,
respectively.  The values of the slopes are shown.  These results
strongly suggest that isolated clouds exist in a different kind of
environment than un-isolated clouds.

Note that the EWDF of \emph{isolated} clouds (Fig. \ref{fig-ewds2},
left-hand plot, dotted line) closely matches the fit to the low EW
part of the mean EWDF.  This convincingly shows that isolated clouds
constitute the ``second'' population (the low-EW side) seen in the
mean EWDF.  As it happens, there are virtually no low EW clouds in the
un-isolated space (see \eg, Fig. 13 in Paper 1).  I parenthetically
note that the line density of void clouds at $N_{HI} \simeq 10^{14}
~\cmtw$ (\ie, about 225 m\AA) is on order 7 per unit redshift.  This
is about $1/5$ of the mean value; certainly not orders of magnitude
less, as implied by simulations of \citet{Riediger:98}.

Henceforth, I shall identify voids as regions that are isolated, where
isolation is determined by the smallness of the scalar value of the
summed tidal field (see \S3 of Paper 1).

The line density of \emph{void} clouds at ${\mathcal W} \ge 15.8$
m\AA\ is $dN/dz \simeq 500$, and the spectral slope is steep
${\mathcal S} \equiv d\,\log( d^2{\mathcal N}/dz \,d{\mathcal
W})/d\log{\mathcal W} \la -1.5 \pm 0.06$ (Paper 1, Table 2).  In
contrast, the slope of the volume-weighted (\ie, mean) EWDF for
${\mathcal W} \ge 32$ m\AA\ is ${\mathcal S} \simeq -0.9 \pm 0.1$,
while non-void clouds have quite flat EWDFs; ${\mathcal S} \simeq
-0.49 \pm 0.12$.  Void clouds appear to actually dominate the mean
EWDF at low EW.

The right-hand plot of Fig. \ref{fig-ewds2} shows the trends of
fitting parameters with changes in limiting values of tide ${\mathcal
T}_{lim}$, for both void (lower sub-panels) and non-void catalogs
(upper sub-panels).  The tide is given in units of inverse Hubble time
squared (see Paper 1) so that values are of order unity.  There is a
relatively narrow transition zone separating clearly distinct
(linearly varying) trends in fitting parameters for void (low
${\mathcal T}$), and non-void (high ${\mathcal T}$) environments (note
the vertical dotted lines).  The transition zone provides a
\emph{phenomenological} basis for establishing the tidal field which
separates void from non-void clouds.  In this way, it is shown that
the space containing void-type clouds fills $\sim 86_{-11}^{+5}\%$ of
the universe (see \citet{Manning:03a} for more details).  This is in
good agreement with the $90\%$ filling factor implied by
\citet{Cen:99}.

I summarize these introductory comments by noting that (1)
N-body/hydro simulations appear to predict that the great majority of
low-$z$ \lya\ clouds are in the diffuse filamentary structure around
galaxies, and not in voids.  (2) The environmentally-resolved
observations of \lya\ clouds in Paper 1 show that most
observed clouds lie in the unshocked void region, even without
corrections for the much smaller redshift coverage of spectra at high
sensitivity\footnote{``High sensitivity'' means sensitivity to very
low EWs.}.  (3) The lack of the ability to resolve cloud halos in
N-body/hydro simulations may be the principal cause for the inability
to predict the existence of this population of void clouds.

The foregoing tentative conclusions already throw suspicion on some of
the results of the simulations.  Is it that simulations fail to
predict void clouds entirely?  Two questions need to be answered;
first, can diffuse clouds, expanding with the Hubble flow, explain the
observed void clouds?  Secondly, if not, can sub-galactic halos
provide the gravitational potential needed to retain enough gas to
produce detectable absorption systems?  This is, in short, the purpose
of this paper.

The paper is organized as follows: In \S\ref{sec-Dave} I investigate
whether diffuse sheets of gas are capable of explaining the observed
void EWDF, as envisioned in simulations.  I find that diffuse sheets
expanding with the Hubble flow are not consistent with observations.
The remainder of the paper assumes the clouds are associated with
sub-galactic halos, and that baryons are subject to pressure gradients
after the reionization of the universe, and tend to expand away from
their halos.  In \S\ref{sec-analytic} I present a general analytical
treatment of spherical clouds to show the relationship between the
baryon density profile and the slope of the column density spectrum.
In \S\ref{sec-lagr} I introduce the hydro/gravity simulation that will
chart the evaporation of baryons from dark matter halos, in response
to the re-ionization of the universe.  These halos are initially
viewed as fully formed and not accreting.  In \S\ref{sec-results} I
present the results of these simulations, based on a range of halo
circular velocities and two halo types.  Section \ref{sec-modeling}
presents a series of strategies for fitting model column density
spectra to the observations of Paper 1.  In \S\ref{sec-extending} I
allow that clouds may accrete sub-halos, and so grow continuously from
$z=6.5$ to the present; the products at $z=0$ are presented, and the
results are compared to those with fixed halos.  In \S\ref{sec-bval} I
reconcile Doppler parameters of model absorbers with those of the
observed void clouds.  In \S\ref{sec-summary} I summarize my findings.

\section{Can diffuse void clouds explain observations?} \label{sec-Dave}

To determine the nature of void clouds under the current paradigm, one
needs to know the fraction of neutral hydrogen atoms.  Assume that
clouds are optically thin, and immersed in a radiation field with an
average flux $J_{\nu}=J_0 (\nu/\nu_0)^{-\alpha_s}$, where $\alpha_s
\sim 1.8$ \citep{Zheng:97, Shull:99} is the spectral index shortward
of the Lyman limit $\nu_0$.  For the local universe, I assume,
\begin{equation} \label{eq-j0}
J_0=1.3 \times 10^{-23} \,\Junits,
\end{equation}
\citep{Shull:99}.  For ionization equilibrium,
\begin{equation} \label{eq-nh1}
n_{HI} =  n_H n_e \frac{\alpha_H^{(A)}}{\Gamma_{HI}}
\end{equation}
\citep{Osterbrock:89}, where $\alpha_H^{(A)}$ is the case-A recombination
rate coefficient appropriate for diffuse clouds.  The photoionization
rate is given by,
\begin{equation}\label{eq-gain}
\Gamma_{HI}= 4 \pi \int_{\nu_0}^{\infty} \left(\frac{J_{\nu}}{h \nu}
\right) \sigma_{\nu} d \, \nu \approx 3.24 \times 10^{-14}
\left(\frac{J_{23}}{1.3} \right) \,{\rm s}^{-1},
\end{equation}
where $J_{23}=J_0/ 10^{-23}$, and $\sigma_{\nu} \simeq
\sigma_0(\nu_0/\nu)^{3}$ represents the hydrogen cross-section for
absorption of a photon of frequency $\nu \le \nu_0$, where
$\sigma_0=6.4 \times 10^{-18} $ \cmtw.

The expected neutral fraction is, from Eq. \ref{eq-nh1},
\begin{equation} \label{eq-nf}
f_{HI}\equiv \frac{n_{HI}}{n_H} = 8.97
\,\left(\frac{1.3}{J_{23}}\right)\left(\frac{\alpha_H(T)^{(A)}}{2.51 \times
10^{-13}} \right)\,n_H.
\end{equation}
Note that in a highly ionized plasma, $n_e \simeq n_H (1+2 \chi)$,
where $\chi = n_{He}/n_H \simeq 0.079$.  Given that $\Omega_m=0.3$,
 $\Omega_b/\Omega_m = 0.1$, and $Y_p=0.24$, the hydrogen
density is $n_H = 5.745 \times 10^{-7} \,h_{75}^2 \Omega_{cl} ~\cmtw$,
where $\Omega_{cl}$ is the total mass density of a cloud in units of
the critical.  Given $\Omega_m=0.3$, the mean hydrogen density is
\begin{equation} \label{eq-nhmean}
{\overline n}_H = 1.725 \times 10^{-7} \,h_{75}^2 ~\cmtw,
\end{equation}
For $T=20,000$ K, $f_{HI}=1.55 \times
10^{-6} ~\,h_{75}^2 \,\Omega_{cl}$.

I note that the time-scale for ionization equilibrium is small;
\begin{equation}
\tau_{eq}^{ion} =
\frac{1}{\Gamma_{HI}} \simeq \frac{1 \times 10^6 \yr}{J_{23}}.
\end{equation}
Thus, ionization equilibrium is a safe assumption.

Now assume that large relatively homogeneous clouds exist in voids, as
generically predicted by simulations, and that they are producing the
isolated absorbers discovered in Paper 1.  Assume that
$\Omega_V/\Omega_m= 0.3$ (where $\Omega_V$ is the total mass density
in voids).  From Eq. \ref{eq-nf}, the neutral fraction is,
\begin{equation} \label{eq-nf2}
f_{HI}=9.38 \times 10^3 \, n_H\, T^{-0.702}\,\h_{75}^2,
\end{equation}
where the index of $T$ is the slope of the recombination coefficient
\citep{Osterbrock:89}. 

\citet{Dave:01} derive a relation between the temperature and the
density from simulated data,
\begin{equation} \label{eq-temp}
T=5000 (\delta\rho/{\overline \rho})^{0.6} \, {\rm K}
\end{equation}
The exponent, 0.6, is actually an increasing function of decreasing
redshift and increases with reionization redshift, but 0.6 appears
consistent with $z=0$ values \citep{Hui:97}.  According to Eq. 4 of
\citet{Dave:01}, $\delta \rho/{\overline \rho} \approx 12
(N_{HI}/10^{14})^{0.7}$ at $z=0$.  The over-density required to
produce a column density $N_{HI}=10^{12.5}\,\cmtw$ is $0.379\,
{\overline n_H}$, or $6.54 \times 10^{-8} ~\h_{75}^2 \,{\rm cm}^{-3}$.
With Eq.  \ref{eq-temp} I derive a temperature $T=2794$ K.
Substituting this into Eq. \ref{eq-nf2}, the neutral fraction of
hydrogen is found to be $f_{HI} = 8.50 \times 10^{-6}$.  Thus the
length of the column needed to produce the absorber is,
\begin{equation}
l = \frac{N_{HI}}{n_{HI}} = \frac{N_{HI}}{f_{HI} n_H} \simeq 1.85 \,{\rm Mpc}.
\end{equation}
If the filament is expanding with the Hubble flow, then the velocity
width, ignoring thermal broadening, would be $\sim 140~\kms$.  Because
the background mass density in voids is expected to be about an order
of magnitude less than in the filamentary medium surrounding galaxies,
the frequency of fluctuations of amplitude $\delta \rho/{\overline
\rho}=0.379$ will be very much smaller in voids than in non-void
regions.

For $N_{HI}=10^{13}~\cmtw$, $\delta \rho/{\overline \rho} \simeq
1.07$, $f_{HI}\simeq 8.23 \times 10^{-6} ~\cmtw$, $T=5207$ K and $l
\simeq 2.14$ Mpc.  The implication is clear; absorbers formed in this
fashion would be seen only as very broad troughs.  The Hubble flow
across $l$ would then be well beyond the values found in the Space
Telescope Infrared Spectrograph Echelle spectra \citep{Dave:01}.  In
GHRS spectra with resolution is $\sim 19$ \kms\ \citep{Penton:00a},
void clouds have \lya\ Doppler parameters, $b \sim 30$ \kms\ (Paper 1),
probably large by a factor of $\sim 2$ due to resolution problems, and
the presence of components \citep{Shull:00}.  Thus void clouds have
Doppler parameters only 1/5 to 1/10 as large as that which diffuse
sheets would produce in voids.

In conclusion, it appears clear that the detected void absorbers must
be either significantly denser than the mean density (\ie,
$\rho/{\overline \rho} \approx 10$), or the velocity field of the
baryons must be strongly convergent with respect to the Hubble flow,
or both.  Neither of these conditions is consistent with the model of
void clouds as diffuse baryonic slabs expanding with the Hubble flow.
Thus, observations appear to be in conflict with the paradigm promoted
by the N-body simulations as applied to \lya\ clouds.

In the following section we begin our analysis of sub-galactic
perturbations as a parent population for void clouds.

\section{An Analytic Treatment of Spherical Clouds} \label{sec-analytic}
If homogeneous slabs of gas expanding with the universe cannot explain
these clouds, then what might?  I hypothesize that absorbers stem from
sub-galactic perturbations that were massive enough to restrain the
outflow of gas from the dispersive tendencies of the Hubble flow and
the pressure gradients that result from the ionizing background. I
start with an analytic treatment of clouds taken as baryons held
within a ``mini-halo'' \cite{Rees:86}.

Clouds whose evaporation is restrained by a centrally condensed halo
of dark matter will have a radial density gradient.  If such clouds
can be roughly approximated by power law density profiles, their
ionization structure, and column density spectra can be calculated.

\subsection{The \hone\ column density of spherical clouds } \label{sec-colden-sph}
I initially assume that the baryons of a cloud have a spherically symmetric
power-law distribution with index $\vartheta$;
\begin{equation} \label{eq-den-profile}
\rho(r) \propto r^{-\vartheta}.
\end{equation}

Eq. \ref{eq-nf} shows that $n_{HI} \propto n_H^2$.  The \hone\
column density produced at projected radius $r_p$ from the cloud
center is,
\begin{equation} \label{eq-colden1}
N_{HI} = \int_{-\infty}^{\infty} n_{HI}(r) \, dl,
\end{equation}
where $l$ is constrained by the equation $r^2=l^2+r_p^2$.  From Eqs. \ref{eq-nf} and \ref{eq-colden1} we find,
\begin{equation}\label{eq-colden2}
N_{HI} =\frac{2 \alpha_H^{(A)}}{\Gamma_{HI}}(1+2 \chi)
\int_0^{\infty} (n_H (r))^2 dl.
\end{equation}

\subsection{The Column Density Spectrum of Spherical Clouds} \label{subsec-CDS}
What column density spectrum would a cosmological distribution of such
clouds produce?  The cumulative column density spectrum may be
broken down by the chain rule,
\begin{equation} \label{eq-diffl}
f(N_{HI})=\frac{{\partial}^2 {\mathcal N}(\ge N_{HI})}{\partial z \,\partial N_{HI}} =
\frac{{\partial}^2 {\mathcal N}}{\partial z \, \partial r}
\frac{\partial r}{\partial N_{HI}},
\end{equation}
where ${\mathcal N} \equiv n_{cl}(\ge N_{HI})$.  The relative number
${\mathcal N}$ of absorption systems produced by a spherically
symmetric system with a column density $\ge N_{HI}$ at projected
radius $r_p$ has a derivative, $\partial {\mathcal N}/ \partial r_p
\propto 2 \pi r_p$.  By Eqs. \ref{eq-nh1} and \ref{eq-den-profile},
$n_{HI} \propto r^{-2 \vartheta}$.  Then by Eq. \ref{eq-colden1},
\begin{equation}
N_{HI} \propto r_p^{-(2 \vartheta -1)}.
\end{equation}
Carrying out the derivatives in Eq. \ref{eq-diffl} and simplifying,
\begin{equation} \label{eq-CDS-diffl}
f(N_{HI})=\frac{\partial}{\partial N_{HI}}\left(\frac{\partial {\mathcal
N}}{\partial z}\right) \propto N_{HI}^{-(2 \vartheta+1)/(2 \vartheta-1)},
\end{equation}
which is the differential distribution.  

The cumulative column density distribution has a slope larger by 1;
\begin{equation} \label{eq-CDS-slope1}
{\mathcal S}=\frac{2}{1-2\vartheta},
\end{equation}
The inversion of this yields,
\begin{equation} \label{eq-CDS-slope2}
\vartheta= -\frac{1}{{\mathcal S}} + \frac{1}{2}.
\end{equation}

These two equations provide an important diagnostic of the observed
column density spectrum (CDS) if we are to interpret it as produced by
discrete, spherically distributed clouds, because it determines the
cloud density profile that is consistent with a given CDS slope.  Note
that the steeper the index of the density profile $\vartheta$, the
flatter the spectral slope ${\mathcal S}$.  For instance, the observed
slope of the void cloud EWDF, ${\mathcal S} \simeq -1.5$ can be
explained if $\vartheta \simeq 7/6$; \ie, $\rho \propto r^{-1.17}$,
while the slope for non void clouds (${\mathcal S} \simeq -0.5$) is
explained if $\rho \propto r^{-2.5}$.

\section{Numerical Simulation of clouds} \label{sec-lagr}

For baryons associated with dark halos, it is not expected that
baryons would follow their halo mass distribution at late times
\citep[e.g.,][]{Barkana:99} -- one might expect a flatter distribution
owing to the effects of evaporation.  As noted above, the steep EWDF
of void clouds (Fig. \ref{fig-ewds2}) imply a relatively flat density
profile; this may be a measure of the effects of reionization on
baryons held by small dark matter (DM) halos.  These considerations
suggest that the clouds producing the absorption in voids have evolved
into a rather flat density profile.  A more accurate view of the
probable profiles of such clouds may be achieved by a cosmological
simulation of the response of DM-held clouds to reionization.

I propose to follow in detail the likely evolution of baryons in
clouds gravitationally held by dark halos before reionization ($z
\simeq 6.5$, \citealp{Djorgovski:01}).  As noted above and in Paper 1,
3-D simulations suffer from space- and mass-resolution problems
when small-scale structures are of interest, making it difficult to
separate actual physical results from numerical artifacts.  When
small-scale structure needs to be studied, a 1-D code may provide a
solution to the problem.  I acquired a copy of a 1-dimensional
Lagrangian hydro/gravity code of \citet{Thoul:95} courtesy of one of
its authors, (A. T.), and adapted it to the needs presented by the
problem posed above.  Among the adaptations made was the
transformation from Einstein-de Sitter universe to a flat FRW
($\Lambda=0.7$) with $h=0.75$, and the writing of initializing
routines for various halo models.  These are consistent with the
parameters used in Paper 1.  A cosmic baryon ratio
$\Omega_b/\Omega_m=0.1$ is assumed.

\subsection{Nature of 1-D Code} \label{sec-nature-code}
The code sets the value of the gravitational constant $G$ to unity.
In addition, it sets a mass scale, ${\widehat M}=1.0 \times
10^{11}~\Msun$, and a length scale ${\widehat L} = 1.0 \times 10^{12}$
AU, or 22.592 kpc, mandating a time scale ${\widehat t}=5.042 \times
10^{15}$ s, or $\sim 160$ Myr.

\subsubsection{Governing physics} \label{sec-Nature-phys}
\begin{figure}[h!] 
\centering
\vspace{-0.5cm}
\epsscale{1} 
\plotone{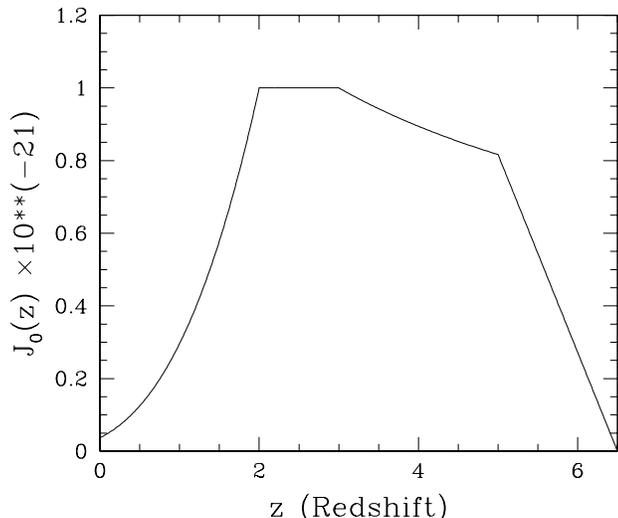}
\vspace{-1.5cm}
\caption{\label{fig-j0}The prescribed average intensity at the Lyman
limit, as adapted from \citet{Thoul:95}.  In the interval $0 \le z \le
2.0$, $J_0(z) \propto J_0(1+z)^3$.  The final results of simulations
are fairly insensitive to the details of reionization.}
\end{figure}

The equations, and related details of the computational scheme that
forms the basis for tracking the evolution of the clouds are detailed
in \citet{Thoul:95}.  The basic components are recapitulated as
follows.  Ionization equilibrium is assumed because its time-scale is
smaller than the time-scales of interest.  The energetic gain $\Gamma$
from photoionization of \hone, \heone\ and \hetwo\ by UV photons is
calculated on the basis of self-consistent concentrations of these
components.  We assume that the spectral index above the Lyman limit
follows the relation, $J_{\nu}=J_0(\nu/\nu_0)^{- \alpha}$, where
$\alpha=1.8$, \citep[\eg,][]{Zheng:97,Shull:99}.  The redshift
variation of $J_0(z)$ is adapted from \citet{Thoul:95}, and is shown
graphically in Fig. \ref{fig-j0}.  The radiative losses, $\Lambda$, are
found by summing the cooling rates from collisional excitation and
ionization, recombination, and bremsstrahlung, given the ionic
abundances and the density \citep{Thoul:95}.  In the Lagrangian code,
bin boundaries for baryons respond to the effects of pressure
differentials and gravitational potential gradients.  The change in
the internal energy per unit mass $u$ with time is then,
\begin{equation}
\frac{du}{dt} = \frac{p}{\rho_g^2} \frac{d \rho_g}{dt} +
\frac{1}{\rho_g}(\Gamma - \Lambda),
\end{equation}
where the first term on the right hand side is the adiabatic loss; $p$
is the pressure, and the $g$-subscript refers to corresponding values
for gas-bins.  Shocks occur when motions are convergent.  The
artificial viscosity technique of \citet{Richtmyer:67} is used,
adjusted to spread shocks over 4 or 5 shells, helping to ensure energy
conservation \citep{Thoul:95}.  The pressure is given by the equation
of state,
\begin{equation}
p = \frac{2}{3} ~\rho_g u. 
\end{equation}
The baryonic component responds to forces as follows,
\begin{equation}
\frac{dv_g}{dt} = -4 \pi r_g^2~ \frac{dp}{dm_g} - \frac{M(r_g)}{r_g^2},
\end{equation}
where $M(r_g)$ includes dark and baryonic mass, $dp$ is the pressure
differential over a Lagrangian bin, and $dm_g$ is the gas mass
contained therein.

The time-step for the simulation is determined by the smallest of a
collection of time scales, including the dynamical timescale, the
Courant, and the cooling time scales, and a time-scale to ensure that
fluid shells do not cross.  In addition there is a
maximum length time-step of $10^{-3}\, {\widehat t}$.

The minimum time-scale is frequently determined by the dynamical
time-scale in the first few shells.  When the timescale for cooling is
much less than the dynamical time scale, shells are essentially in
free-fall.  This causes the central density to increase dramatically
so that cooling time steps become extremely small.  In response to
this problem, when $\tau_{cool} \ll \tau_{dyn}$ the program
``freezes'' the shells in question, and refers the mass to a minimum
radius, here, $r_{min}=r_c$ -- essentially taking the mass in the
shell out of gas form (we might speculatively suggest that gas
fragments and forms stars in the collapse zone).  In practice, this
does not occur for systems of circular velocity less than $v_c \la 21$
or 24 \kms, for \citet{Navarro:96, Navarro:97} (hereafter, NFW) and
isothermal halos, respectively.  Since this is an investigation of
clouds and not galaxies, I ignore systems of larger circular velocity.
I check that the point at which shells are frozen is physically
realistic by monitoring the ratio of the diameter of the first
Lagrangian bin to the Jeans length, given by
\begin{equation}
\lambda_J=\sqrt{\frac{\pi \gamma k_B T}{G \rho \mu  m_H}},
\end{equation}
where $\gamma$ is taken to be $5/3$, and $\mu$ is the mass per
particle in units of the hydrogen atom, that approaches the value
$\mu=0.588$ at the low densities of model clouds at low to moderate
redshifts.  This indicator is found to approach unity shortly before
the shells are frozen, indicating that the program is treating the
problem correctly.

It is possible that starting with a less-relaxed halo, warmed by
secondary infall at $z < 6.5$, halos of circular velocity high enough
to collapse in this way under the current arrangement, would not
collapse.  This implies that clouds of somewhat larger circular
velocity could exist without forming stars (see \S\ref{sec-extending}).

\subsubsection{General Treatment of Dark Halos }
Because the \citet{Thoul:95} code was originally written for
collapsing protogalactic clouds, the program's treatment of DM is
inappropriate for the present simulation.  Instead, a dark halo
overdensity is established and not allowed to change (these halos are
to be specified below).  It would be more proper to project a dark
halo as it would be at a high redshift, growing somewhat by secondary
infall, so that the end product is consistent with the specified
circular velocity.  However, of this there are two views; one states
that the halo of a small cloud is in place by $z=6.5$ (see
\citealt{Wechsler:02})\footnote{The virialized mass will grow in time,
but that is only apparent growth since it results from the decline in
$\rho_{crit}$ with time.}, so that most of the response to absorbed
radiant energy is at redshifts in which the total cloud mass is fairly
stable.  In this scenario, therefore, assuming a fixed halo circular
velocity can only modestly minimize the final extent of cloud
expansion due to its greater average potential.  However, the other
view is based on observations suggesting the growth of halos, even at
late times.  Therefore, after applying the analyses to fixed halos, we
then allow halos to grow steadily by accretion of sub-halos, leading
to a growth of circular velocity with time.  These ``grown'' halos are
subjected to the same analysis as the fixed halos, providing a
plausible maximal range for reasonable models of halo growth.

The spherically symmetric conditions of the simulation presume that
the halos are isolated from the influences of others.  This
approximation should be accurate for void clouds since the environment
around them is expanding with the Hubble flow, and evaporation ejecta
are not likely to interact with that of other isolated systems.
However, it would not likely apply to non-void space because in the
convergent flows of over-dense regions, cloud outflows may collide
causing shocks, that violate the isolation and spherical symmetry
assumed in this simulation.

\subsubsection{General Initial Conditions}
The initial conditions of model clouds are designed to make them
dynamically stable.  Because the central densities of clouds are
supposed to approximate the density of the universe at the redshift of
their formation (the canonical explanation for the high concentration
parameters of small systems in relation to those of large systems
found in simulated halos), I assume that the halos I model are already
in place at $z=6.5$.  There may be some confusion on this issue
because it is seen that the virial mass of even small halos continues
to grow down to $z=0$ \citep[\eg][]{Bullock:01}.  However, the virial
radius is approximately equal to the radius within which the average
density is $\sim200~\rho_{crit}$.  Therefore, the decline in the
background density with the expansion of the universe causes an
expansion of the virial radius, resulting in an increase in the virial
mass with time.  Comparisons with code kindly lent from J. Bullock
(``cvir.f'') show that, for clouds with $v_c \le 25 $ \kms, there is
very little increase in mass, or decrease in the concentration
parameter over and above that which naturally results from the
expansion of the universe.  Therefore, from the standpoint of current
thinking, assuming that the halos were in place at $z=6.5$ should be
non-controversial.  However, from the standpoint of the isothermal
model, this may not be true.  Accordingly, we do investigate halos
which grow steadily in \S\ref{sec-extending}.

In deciding the initial conditions of the gas associated with a
``fixed'' dark halo (\ie, fixed circular velocity $v_c$ and
overdensity profile), it is important to note that at high redshift
the baryonic cloud is immersed in a critical density background with
density comparable to that of the cloud.  For, at z=4.84 the
background density is 200 times the current critical density, thereby
making the virial radius significantly smaller than at the present.
Because baryons are dissipational, there is no way to separate the
baryons that are part of the overdensity from those that are part of
the background.  In fact, what allows for the existence of a prolonged
and stable overdensity is a distortion in the Hubble flow that is
concentric about a point we would call the center of the cloud.  Since
these halos are immersed in a high-redshift universe, their
overdensity exists by virtue of this deviation from the Hubble flow.

At the initialization stage, an outward motion is given to the gas
that is less than, but gradually approaches the Hubble flow velocity
as one considers radii farther from the cloud center in a manner such
that the overdensity is stable in time.  This is actually easier than
it sounds, for one need only give an outward velocity such that the
radial flux of gas is what it would be if the universe were flat and
of critical density.  Thus, at a given redshift $z$, the outward
motion of gaseous matter $v_g$ needed to maintain a density
$\rho_g(r)$ at a distance $r$, for instance, should equal,
\begin{equation}
v_g(i)=H \, r(i)\, \Omega_b
\,{\rho_{crit}}/\rho_g(i),
\end{equation}
where the variables, $H$, $\Omega_b$, and $\rho_{crit}$ are evaluated
at the redshift in question.  This creates an overdensity that is
stable within the Hubble flow (the peculiar velocity supports the
overdensity).  The dark matter is given the same peculiar velocity
field.  The velocity field that results can be expressed as proper
radial motion (including the Hubble flow) as well as peculiar
velocity, with the Hubble flow subtracted.

Because of the exigencies of a Lagrangian code, and the apparent lack
of evidence for a truncation of halos at the virial radius, it was
decided not to impose a radial truncation of the overdensity of halo
models.  The halos were extended to a distance at which the average
internal \emph{over}density in units of the current critical density
was ${\bar \rho} = 2 ~\rho_{crit}$, equivalent to a distance of $\sim
2$ Mpc from an ${\mathcal L}^*$ galaxy with an isothermal (hereafter
referred to as ISOT) model halo.  Later, when estimates of the average
density contributed by clouds are required, it will be possible to
impose a truncation radius, such as that featured in Paper 1.  It
should be borne in mind, however, that there are no features in the
simulation products that would imply that the mass is truncated at any
characteristic radius, nor does the preponderance of observational
evidence (see below) require it; only N-body simulations suggest
something like that might be required.  Since the extent of gravity is
infinite, halos have the tendency to draw matter toward them
incrementally at great distances.

Initially, baryons are taken to ``follow'' the DM, so there is an
inward peculiar gravity acting on the baryons.  To restrain baryon
collapse, I initially constrain the cloud to have a temperature such
that the pressure exactly balances the peculiar gravity in each bin.
This will not guarantee that clouds are static because cooling rates
vary within the cloud, however, it is dynamically stable.  For
instance, if the simulation is started at $z=10$, instead of 6.5, the
baryons will ``slosh'' about within their halo, but when ionization
begins, the cloud expands, producing a result remarkably similar to
those with simulations begun at $z=6.5$.

It turns out that the details of the temperature structure is
relatively unimportant because, for clouds with virial temperature
$T_{vir} \le few \times 10^4$ K, the cooling time-scale is large,
suggesting that clouds are fairly stable before reionization.
Further, the energy deposited upon reionization dwarfs the initial
energy content, so that the effect of the initial temperature on cloud
dynamics is insignificant.  Other tests showed that the final state of
the cloud is also quite insensitive to variations in the specific
prescription for the onset of the ionizing flux (see
Fig. \ref{fig-j0}).  Thus, final results are sensitive only to halo
type and circular velocity.

\subsubsection{Initializing routines} \label{sec-nature-init}
In the initialization stage of the program, the halo model and its DM
circular velocity are specified, and the total mass of the cloud in DM
and baryons derived therefrom are calculated according to the cloud
halo model, extending to a radius at which the internal over-density
is 2 times the critical.  This ensures a large enough scale to capture
the relevant dynamics of expansion, and guarantee that all observable
column densities occur within the cloud end-products.  The mass is
then divided equally into the number of Lagrangian bins to be used
(generally, the number of gas bins are set to $n_g=200$, and the dark
matter divided into $n_d=500$ bins).  Then the initial radius at which
these masses are found is determined by straightforward application of
model relations (see \S5.1). 

This study considers halos which are isothermal as well as NFW halos.
I do not truncate these halos at the virial radius because what is of
interest is the mass distribution of the halos, rather their kinetic
properties.  Numerous studies now support the claim that there is no
sign of a drop in the index of the density profile beyond the virial
radius, or for that matter, the NFW radius of maximum circular
velocity \citep[\eg,][]{Zaritsky:94, Zaritsky:97a, Zabludoff:00,
McKay:02}.  In their study of weak lensing, \citet{McKay:02} show that
the line-of-sight (LOS) velocity dispersion $\sigma_r$ of satellite
galaxies around 618 isolated host galaxies from the Sloan Digital Sky
Survey is consistent with $\sigma_r$ being constant over a range $100
\la r \la 500 ~\h^{-1} ~\kpc$, and that the number density of
satellites $n \propto r^{-2.1}$ over this range.  This is very
suggestive of a mass distribution that is not truncated; one not
inconsistent with an isothermal mass distribution.  To make the
distinction clear, when I refer to a halo, I am referring to the mass
distribution, rather than the mass in the virialized portion thereof.
It is expected that the mass of the early halo will generate peculiar
velocities in the surrounding universe, contributing to an enhancement
of the density outside the virialized region of the halo.

{The NFW Halo:} The NFW halo is treated in considerable detail in
Paper 1.  Equations 34 to 40 of Paper 1 describe the derivation of the
scaling relations of the model, which are necessary for calculating
halo parameters as a function of halo circular velocity.  A formula
for the mass as a function of distance is the basis for determining
the bin distributions and calculating density within the bins,
\begin{equation}
M_{\rm NFW}(r)= \frac {4 \pi}{3}{200 \rho_{crit} \,
r_{200}^3}\left\{\frac{\log_e(1+r/r_s)-
\frac{r/r_s}{1+r/r_s}}{\left(\log_e(1+c)-c/(1+c)\right)}\right\},
\end{equation}
where $r_s$ is the scale length, $r_{200}$ is the assumed virial
radius enclosing a density of $200 \rho_{crit}$, and $c$ is the
concentration parameter defined by, $c \equiv r_{200}/r_s$, where
$r_s$ is the scaling radius, and $r_{200}$ is the density at which the
average internal density is 200 times the critical density, or,
loosely, the virial radius.  The concentration parameter is a function
of the halo circular velocity, and was derived from relations taken
from NFW (see \S5.1.2 of Paper 1),
\begin{equation}
c \simeq 132.3 \,v_{max}^{-0.393},
\end{equation}
where $v_{max}$ is the maximum halo velocity, which occurs at a
distance, $r_{max} \simeq 2.16 r_s$.  Be aware that there is significant
dispersion in these relations, as they are average values derived from
the results of computer simulations.  The NFW halo has a cuspy core,
with density approaching $\rho \propto r^{-1}$ as $r$ goes to zero.

{The Isothermal Halo:} In Paper 1, the isothermal halo (which I
will henceforth abbreviate as ISOT) was used to model the total mass
of galaxies, and so required no internal structure to be stated.  This
study, however, requires an explicit core.  I assume that the central
non-singular core of the ISOT model has a density equal to that found
by \citet{Firmani:00} in their investigation of the central densities
of dwarf galaxies.  I apply an exponential term to the density law,
\begin{equation}
\rho(r)=\frac{G {\mathcal K}}{4 \pi r^2}\left(1-e^{-(r/r_0)^2}\right),
\end{equation}
so that that there is a flat core of density $\rho_c \simeq 0.02
~\Msun ~\pc^{-3}$, that constrains the core size
$r_0$, and also allows the mass to approach a value $M(r)={\mathcal K}
r$ for $r\gg r_0$.  I find that
\begin{equation}
r_0=0.3035 \times \, \left( \frac{v_c}{10 ~\kms} \right) ~\kpc.
\end{equation}

The simulation begins with the baryons ``following'' the DM
distribution, and shortly thereafter they are subjected to an
isotropic metagalactic flux whose amplitude at the Lyman limit
$J_0(z)$ is shown in Fig. \ref{fig-j0}.

\begin{figure}[h!] 
\centering 
\epsscale{1} 
\vspace{-0.5cm}
\plotone{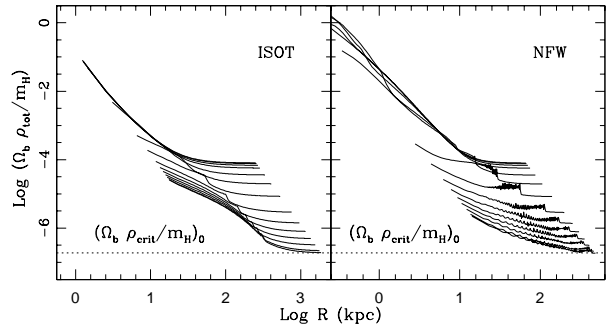}
\vspace{-3.7cm}
\caption{\label{fig-lagr-den2} The evolution of an isothermal cloud
(left panel), and an NFW cloud (right panel), both with $v_c=21.1$
\kms.  The log of the baryon number density is shown as a function of
the log of the cloud-centric radius at an arbitrary series of times
(see text).  The simulation starts at $z=6.5$ (topmost lines) and
proceeds to $z=0$.  The NFW halo (right-hand plot), being more
strongly concentrated, has a smaller first gas bin, and higher density
there, but a lower total mass.  Note the outflowing wave that is
stronger in the NFW halo than the ISOT.}
\end{figure}
\begin{figure}[h!] 
\centering
\epsscale{1} 
\plotone{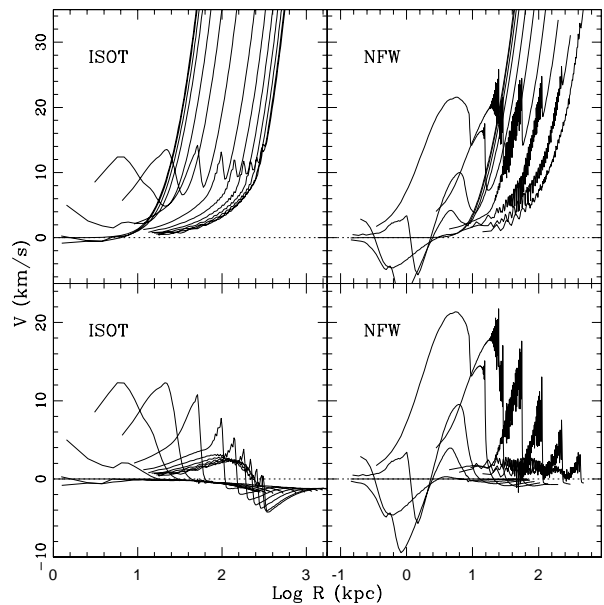}
\caption{\label{fig-lagr-vel4}The proper velocity (upper panels) and
the peculiar velocity (lower panels) for ISOT halos (left panels) and
NFW halos (right panels) at arbitrary points from $z=6.5$ (left-most
line) to $z=0$ (right).  As in Fig. \ref{fig-lagr-den2}, velocities
are shown as a function of the log of the system-centric radius.  The
cloud halo velocity shown in this experiment is $v_c=21.1$ \kms.  This
NFW halo has acoustic oscillations on time-scales of order 600 Myr.
The ISOT model produces relatively minor shocks and oscillations.}
\end{figure}

\subsection{Behavior of Baryon Clouds after Reionization}
The program advances in steps and needs about 15,000 time steps to get
to $z=0$.  The program prints out data every 20 steps, leaving about
separate 750 files.  Figures \ref{fig-lagr-den2} and
\ref{fig-lagr-vel4} show the reaction of a large cloud ($v_c \simeq
21.1$ \kms) to reionization.  The time steps at which the
distributions were sampled are arbitrary; chosen to give a sense of
the dynamics and to avoid confusion.  The chosen files are roughly
geometrically arranged (\eg, using files \#1, 2, 5, 10,..) until
(for ISOT) reaching file number one hundred, then continued to sample
at each 100 steps to the last step near 750.  For NFW halos the
doubling continues to $z=0$ to avoid the confusion of overlapping
plots.  A sense of the redshift can be had by recognizing that the
density at the extreme edge of the cloud is $\Omega_m \rho_{crit}/m_H
\propto (1+z)^3$.

The principal distinction between the ISOT and NFW density profile
evolution is that ISOT has a softer outflowing wave, and a
significantly larger central density at the conclusion of the run.
The NFW halo, which is very concentrated, sends out a strong wave of
expansion that introduces a shock that reverberates back into the
inner cloud.  The lower final central baryon densities of the NFW
clouds, roughly a tenth of the ISOT clouds when comparing clouds of
equal circular velocity, reflects its lower mass.  The evolution of
velocity profiles for halo models with $v_c=21 $ \kms\ is shown in
Fig. \ref{fig-lagr-vel4}.  It shows that the outflowing wave from the
NFW halo has a significantly stronger amplitude -- by almost a factor
of two -- accompanied by a strong oscillation of the inner cloud.  The
magnitude of the wave declines significantly in time for each model.
The final peculiar velocities (lower panels) at large cloud-centric
distances are only slightly negative for the NFW halo, while the more
massive ISOT halo has produced a sizable negative peculiar velocity at
late times just beyond the outflowing wave.  Note, however, that the
higher density gas internal to this still has a positive peculiar
velocity.  This discontinuity, here located at $r \simeq 260 ~\kpc$,
is far beyond the canonical virial radius for this cloud ($r_{vir}
\simeq 25 ~\kpc$).

\subsection{Comparison of Program Output with Other Work} \label{sec-Hui-cmpr}
\begin{figure}[h!] 
\centering
\epsscale{1.0} 
\plotone{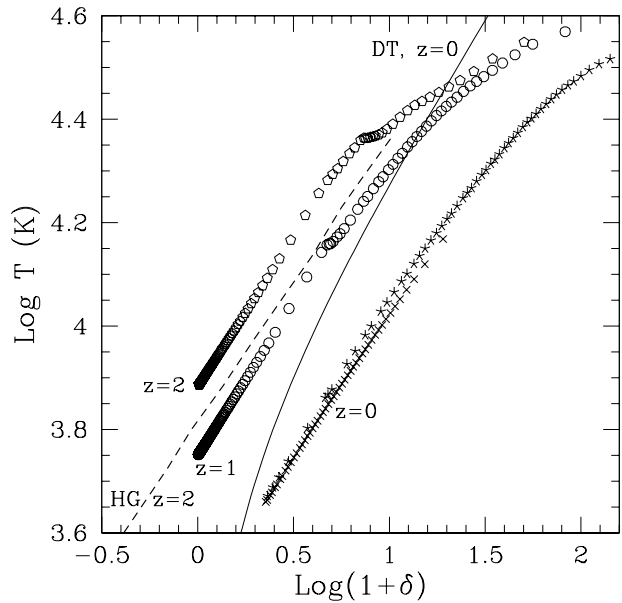} 
\caption{\label{fig-lagr-hui} The logarithm of the temperature (in
Kelvins) of cloud parcels as a function of the log of the over-density
(in units of $\Omega_b \, \rho_{crit}$) for various redshifts.
Symbols represent the present work, and labeled lines represent the
work of \citet{Hui:97}, and \citet{Dave:01}.  The skeletal symbols
show the variation of trends of temperature at $z=0$ in large clouds
(5-armed) vs. small clouds (4-armed), where smaller clouds are cooler
as a result of the greater freedom a small cloud has to expand, and
adiabatically cool.  Open circles, and pentagons show trends for $z=1
$, and 2, respectively, for the large ($v_c=23.7 $ \kms) cloud.  See
\S\ref{sec-Hui-cmpr} for further discussion.}
\end{figure}

Data from simulations of clouds can be used to probe the trend of
temperature with density as a function of redshift.  In
Fig. \ref{fig-lagr-hui} is plotted the logarithm of the density (in
units of $\Omega_b ~\rho_{crit}$) \vs the log of the temperature at
z=0 (skeletal), $z=1$ (open circle), and 2 (open pentagon).  The data
at z=0 is taken from a maximal ISOT halo with $v_c=23.7$ \kms\
(5-pointed stars) and with a smaller halo, $v_c=13.3 $ \kms\
(crosses). Note that the locus of points from the smaller cloud is
marginally cooler than that from the larger cloud because the smaller
cloud may expand more freely, increasing the adiabatic cooling.  The
heating produced by the outflowing shock (or cooling following its
passage) can be readily seen in the $z=1$ and 2 plots (note the
discontinuity), though by $z=0$ the wave has passed beyond the
outermost point.

Also in Fig. \ref{fig-lagr-hui}, these program outputs are compared
with a result from Fig. 3$b$ of \citet{Hui:97}, based on the output of
a 3-D hydro code (dashed line, label $HG$, $z=2$), which utilizes
rapid reionization at $z=7$ and a flat $\Lambda$ model.  Also shown is
the relation implied by Eq. \ref{eq-temp}, as found in \citet{Dave:01}
(solid line, labeled $DT$), which represents the trends in the
filamentary structure surrounding galaxies (\ie, non-void space) at $z
\simeq 0$.  Note that the \citet{Hui:97} trends ($z=2$) appear to
approach the trend for the results from the 1-D simulation
\emph{inside} the outflowing wave.  However, the \citet{Dave:01} model
appears to predict significantly larger temperatures at densities of a
few or more.  This is attributed to shock heating and bulk motions in
this denser environment, said to be more pronounced at low redshift
\citet{Dave:01}.

Given the differing physical conditions that are probed by the 3-D
(non-void) and the 1-D (void) simulations, the differences in
predicted temperatures are not surprising.

\section{Results at Redshift Zero} \label{sec-results}

The program is run for a range of cloud halo velocities $v_c$ with
bin-centers at,
\begin{equation} \label{eq-bin-cntr1}
v_c(n)=5.31 \times 10^{0.05 \, n} ~\kms,
\end{equation}
for n=0 to 13.  These bins cover circular halo velocities of 5 to 25
\kms.  Halos larger than this experience what appears to be the
equivalent of star formation, \ie, free-fall collapse in the central
regions of the cloud.  The cadence in Eq. \ref{eq-bin-cntr1} has 20
logarithmic steps in a decade of $v_c$.

Another form for designating the velocity bins is the following,
\begin{equation} \label{eq-bin-cntr2}
X_v(n)\equiv\frac{v_c(n)}{v_c^*} = \frac{v_c(n)}{144}.
\end{equation}
One should notice that 144 \kms\ is the velocity of the dark halo of
an ${\mathcal L}^*$ galaxy that, when baryons are added to comprise 10\%
of the total, has a circular velocity of 161 \kms\ (see \S5.1.1 of
Paper 1).

\begin{figure}[h!] 
\centering
\epsscale{1.0} 
\plotone{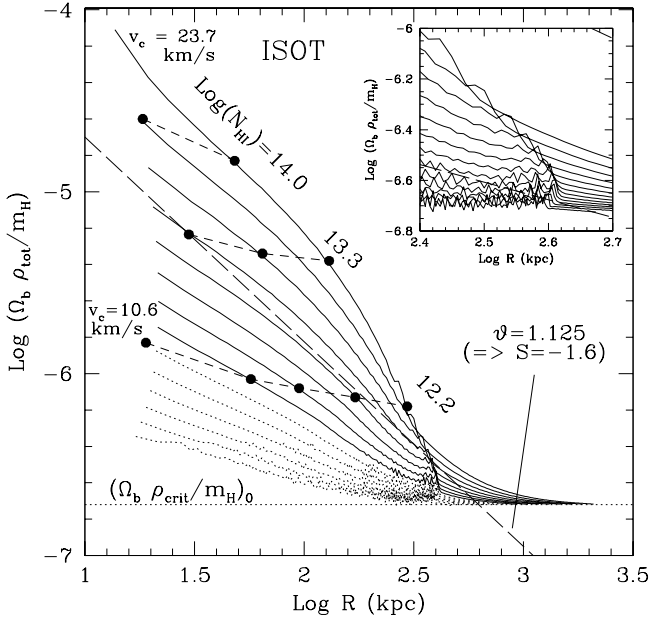} 
\caption{\label{fig-deniso}Final density profiles ($z=0$) for the
range of halo velocities for the ISOT model.  The dashed line
represents the slope of the density profile ($\vartheta=1.125$)
necessary to produce the observed slope ${\mathcal S}\approx-1.6$ of the
CDS for void clouds, (see \S\ref{sec-colden-sph}).  Clouds with
circular velocity $v_c \ga 10 $ \kms\ have slopes that approximate this
value inside $\sim 150$ \kpc.  Profiles with dotted lines are
undetectable in \citet{Penton:00a}.  The enlarged dots and
short-dashed lines represent the radii where particular clouds give
the stated column density (see log of column density labels).  The
inset shows the detail of the cloud outer edge showing that the
self-gravity of the larger clouds has restrained the outflow wave for
halos with velocity $v_c \ga 10 $ \kms. }
\end{figure}

\subsection{Model Halos at $z=0$}
The products of simulations for $z=0$ are of special interest, for the
density profiles, temperatures, and neutral fractions of hydrogen,
will provide what we need to produce a model column density spectrum
(CDS).  The results strongly depend on the halo model.

\subsubsection{ISOT density profiles}\label{sec-ISOT-profiles}
The $z=0$ results of an ISOT simulation are seen in
Fig. \ref{fig-deniso}, where the smallest cloud $v_c=5.31 $ \kms\ is
the bottom, dotted line.  The smallest cloud to produce a detectable
absorption line of $N_{HI} \ge 10^{12.4} ~\cmtw$ is the bottom solid
line, where$v_c=10.6$ \kms.  The long-dashed line in the figure has a
slope $\vartheta$ that is associated with a CDS slope ${\mathcal S} =
-1.6$ through Eq. \ref{eq-CDS-slope1}, the observed slope of the void
CDS.  The labeled short-dashed lines show the radius at which selected
column densities are detected, and the baryon density at the closest
approach of the LOS.  This will be discussed in detail in
\S\ref{sec-EW-colden}.  Note that large clouds may produce low-column
density absorbers at distances as large as 200 or 300 kpc, though the
spectral slope they would produce would be significantly shallower
than that observed in the void CDS of Fig. \ref{fig-ewds2}.

The inset of Fig. \ref{fig-deniso} shows an enlargement of the zone
of the outflowing wave.  For smaller halo velocities, the position of
the wave at $z=0$ scales directly with the core-radius, indicating the
outflow is essentially unrestrained.  However for $v_c \ga 10.6 \,
\kms$ (8th from the top), the position of the wave is at progressively
smaller radii, attributable to the cumulative effects of their massive
halos.  These effects are also seen in the peculiar velocity plot of
the large ISOT halo in Fig. \ref{fig-lagr-vel4}.

That neutral hydrogen mass of these halos can be straightforwardly
calculated from the output of the code, and while not directly
relevant to this study, it is relevant to 21 cm studies of high
velocity clouds (HVCs), and is an opportunity to compare \lya\ and
HVCs.  These data are presented in \S\ref{sec-extending}.

\subsubsection{NFW density profiles} \label{sec-NFW-profiles} 
\begin{figure}[h!] 
\centering
\epsscale{1.0} 
\plotone{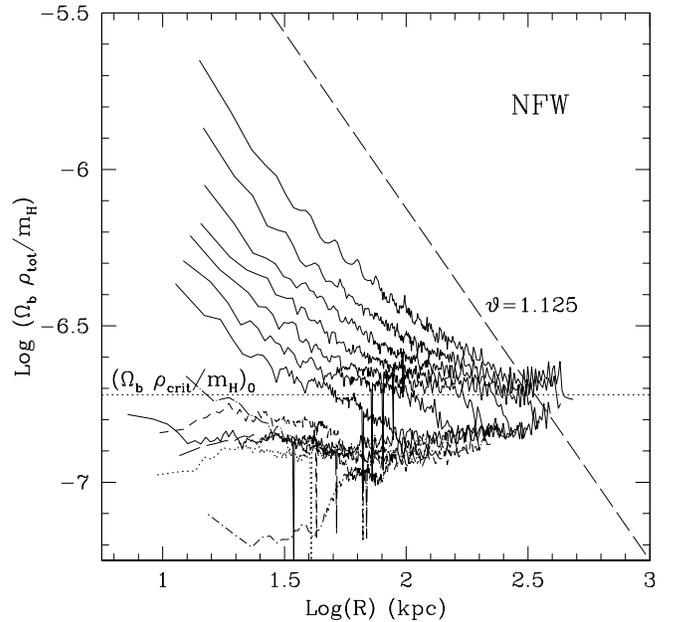}
\caption{\label{fig-den-NFW} NFW density profiles evolved to $z=0$ for
$v_c=5.3$ to 23.7 \kms.  Note that low $v_c$ clouds have expanded with
such force that they are actually \emph{under}densities at $z=0$.  The
dashed line has slope $\vartheta=1.125$, that would produce a column
density spectrum slope ${\mathcal S} = -1.6$, approximately that of the
void CDS.}
\end{figure}

The NFW simulations to $z=0$ are presented in Fig. \ref{fig-den-NFW}.
Gas density profiles are seen to be flatter than $\vartheta=1.125$
outside $\sim 30 \, \kpc$.  There is an apparent total blow-out
phenomenon for clouds with $v_{max} \le 10 \,\kms$, and a blowout of
the outer cloud for clouds smaller than $\sim 18 $ \kms.  The smaller
physical extent of the clouds has resulted in the expanding wave
having traveled beyond the outermost Lagrangian bin except for the
most massive cloud with $v_{c}=21.1 \,\kms$.  The cloud
with $v_c=23.7 $ \kms, however, collapsed from the inside out, and
would presumably form stars, and was therefore discarded as a model
cloud.  As we shall see, NFW halos do not produce observable \hone\
columns in these simulations.


\subsection{The Nature of Model Absorbers} \label{sec-model-abs}
Here I discuss the method of extracting model absorption systems,
and looking at the thermal and
velocity structure of the absorbers.

\subsubsection{Calculating cloud column density} \label{sec-rp-colden2}

Let us consider a line of sight through a particular ISOT cloud.
Given an arbitrary targeted column density, we seek to find two
consecutive bin radii that serve as impact parameters $r_p(i)$ and
$r_p(i+1)$, at which column densities bracket the target column.  One
may then interpolate the solution.  When the line of sight (LOS)
velocity profiles of the model absorbers are analyzed, it is found
that absorption systems typically have a strongly peaked central core,
with broad wings that may add substantially to the calculated column
density.  This ``background'' is attributable to a Gunn-Peterson
effect \citep{Gunn:65}.  Simulated clouds are hypothesized to be
immersed in a gas of density $\Omega_b\rho_{crit}$, and under the UV
flux utilized in these simulations (see Fig. \ref{fig-j0} and
Eq. \ref{eq-j0}), the neutral fraction of hydrogen gas of this density
at $z=0$ is found to be $f_{HI} \simeq 1.5 \times 10^{-6} ~\cmthr$.

The simulated data is subjected to the same treatment as the
observations.  In the reduction of the observational data, a continuum
is placed with respect to which the fluctuations of the spectrum, in
zones not influenced by absorption or emission lines, are evenly
balanced, above and below the continuum.  Therefore, this background
Gunn-Peterson effect is subtracted from the model absorber.  Using the
neutral fraction of this background gas, and the $z=0$ background
hydrogen density ($1.745 \times 10^{-7}~\,{\rm cm}^{-3}$), yields a
neutral hydrogen density in a cloud of total density $\Omega_m
\rho_{crit}$ of $n_{HI}=2.61 \times 10^{-13}$ \cmthr.  The neutral
column in a megaparsec is thus $N_{HI}=8.04 \times 10^{11} $ \cmtw.
In the Hubble flow, if we take 1 Mpc to be 75 ~\kms, there is a column
``dispersion'' of $d \,N_{HI}/dv=1.07 \times 10^{10} \cmtw$ per \kms,
or $10^{12} $ \cmtw\ per 93.3 ~\kms.  This provides a minimal
approximation to the displacement of the continuum due to background
neutral hydrogen.  This background column density dispersion is scaled
to the velocity of the extracted profile, and subtracted from the
total column density to get an ``observed'' value.


The program finds the projected radius $r_p$ at which a target column
density occurs, and uses that to calculate the surface area of the
cloud that will produce a column greater or equal to the target
column.  The utility of this parameter for deriving the column density
spectrum will be demonstrated in \S\ref{sec-CDS}.

\subsubsection{Cloud temperature versus $b$-value} \label{sec-T-b}
The temperature of a cloud is an important factor in predicting the
Doppler parameters of model clouds.  As shown in
Fig. \ref{fig-lagr-hui}, the temperature of clouds vary
systematically with density.  I define the temperature of an
absorption line as the temperature of its core.  To do this I
calculate the column density-weighted temperature within the range at
which the differential column in the successive bins drops to half the
maximum value.

Cloud outflows at $z=0$ are generally quite small (of order $\sim 2$
to 7 \kms) compared to the observed $b$-values of clouds.  Recall that
the most likely Doppler parameter for void clouds with the GHRS data
of \citet{Penton:00a} (see Paper 1, Fig. 6) is $b_{Ly \alpha}\simeq
30-35$ \kms.  If purely thermal, this would correspond to a
temperature of $T\simeq 70,000$ K.  I retrieve from the model lines a
column density-weighted temperature for a range of columns $N_{HI}$,
and a range of cloud velocities, all summed within the line FWHM.
This will provide a theoretical lower limit to observed cloud b-values
as a function of EW or column density.  In fact, a fairly tight
relationship between column density and temperature results from model
clouds.  These temperatures, for log-column densities of 12.5, 13.0,
13.5, and 14.0 are, 7800 K, 11600 K, 16900 K, and 27000 K,
respectively.  The range of temperatures at log-column 12.5 is fairly
large ($\sim 300$ K) because of a wide range of halo velocities that
may produce the line, but it is smaller for large column densities
since only a small range in cloud $v_c$ may produce them.  These data,
transformed into a thermal $b$-value\footnote{for hydrogen,
$b_{therm}\simeq 0.128 \, T^{1/2}$ \kms}, are plotted in
Fig. \ref{fig-bscatt} as open pentagons, and provide a theoretical
lower limit to cloud $b$-values..

\subsection{Transforming EW into Column Density} \label{sec-EW-colden}

In order for observed absorption systems (EW in units of m\AA) to be
compared with model results (column density in units of \cmtw), one
needs to account for the self-blanketing effects of \hone\ clouds.  In
this section I discuss transformations of the EWDF to a CDS, for this is
required before model clouds may be compared to the observed EWDF.

If the cloud is quiescent and of uniform temperature, then the thermal
broadening of the line will characterize the Doppler parameter $b$.  A
low-temperature cloud will deplete the continuum photons with
wavelengths $\lambda=1215.67$ \AA\ in the cloud rest frame because of
the large fraction of neutral atoms that can efficiently absorb at the
\lya\ resonance, and so produce an absorption system with lower EW
than a hotter cloud of the same column density, which would not
produce this ``blanketing''.  If there is well-developed turbulence
such that at physical scales of the overturning regions, the column
density is small in relation to the total column through the cloud,
then this must be added in quadrature to the thermal $b$-value in
order to arrive at the total $b$-value; $b=\sqrt{b_{therm}^2 +
b_{turb}^2}$.  If turbulence or virial motions exist on large scales,
then the cloud should ideally be characterized by a superposition of
independent Doppler parameters, rather than a single one.

The measurement of $b$ from the \lya\ line of an \hone\ absorber
(hereafter $b_{Ly \alpha}$) may reflect not only the line broadening
effects of temperature and small-scale turbulence, but large-scale
cloud motions, or physically associated sub-condensations; thus it is
also a function of the resolution of the spectra.

For clouds with no large-scale motions, the line-blanketing effects in
clouds can be characterized by the use of the relation of
\citet{Chernomordik:93} that approximates the equivalent width of
an absorber as a function of optical depth $\tau$ and the Doppler
parameter $b$;
\begin{displaymath}
{\mathcal W} = \frac{\sqrt{2} b
\lambda}{c}\left\{\log_e\left(1+\frac{\pi}{2}\tau^2\right)
\right\}^{1/2}.
\end{displaymath} 
The optical depth $\tau = 1.497 \times 10^{-2} \lambda f_{12} N/b$,
where the oscillator strength for \lya\ is $f_{12}=0.416$, and
$\lambda=1215.67$ \AA.  Substituting, and solving for column density,
\begin{equation} \label{eq-cherno}
N_{HI} = 1.05 \times 10^{12} \,b \sqrt{\left\{\exp\left[\frac{1}{2}
\left(\frac{0.247 \, {\mathcal W}}{b}\right)^2\right]-1\right\}} ~\cmtw.
\end{equation}
This relation is claimed to be accurate within 10\% for all optical
depths $\tau < 40$.  For small ${\mathcal W}$ (${\mathcal W} \la 150$
m\AA), it reduces to the standard linear relation,
\begin{equation}
N_{HI} = \left(\frac{\mathcal W}{54.43}\right)\times 10^{13} ~\cmtw,
\end{equation}
as noted in \citet{Penton:00b}. 

It is possible to separate the effects of the large scale motions
influencing the \lya\ line-width from the effects properly
incorporated within the Doppler parameter by using observations of
higher-order Lyman series absorbers and curve-of-growth analysis to
arrive at the true $b$-value.  Using the Far Ultraviolet Spectroscopic
Explorer (FUSE), \citet{Shull:00} observed 12 clouds with ${\mathcal
W} \ga 200$ m\AA.  Results suggested that the intrinsic (thermal) line
widths of the clouds were generally much smaller than that measured
from the \lya\ line.  For the clouds with modest to small
uncertainties in b-value, the average reduction from $b_{Ly \alpha}$
to $b$ was a factor of from 0.5 to 0.57, with dispersions in the range
$\sigma=$ 0.15 to 0.19.  In studying Shull's results, the fractional
reduction of $b$-value showed no apparent correlation with cloud EW.
The large dispersions indicate that there is a significant random
element in the fraction of $b_{Ly \alpha}$ that is due to bulk motions
or clustered sub-condensations.

\begin{figure}[h!]  
\centering
\epsscale{1.0} 
\plotone{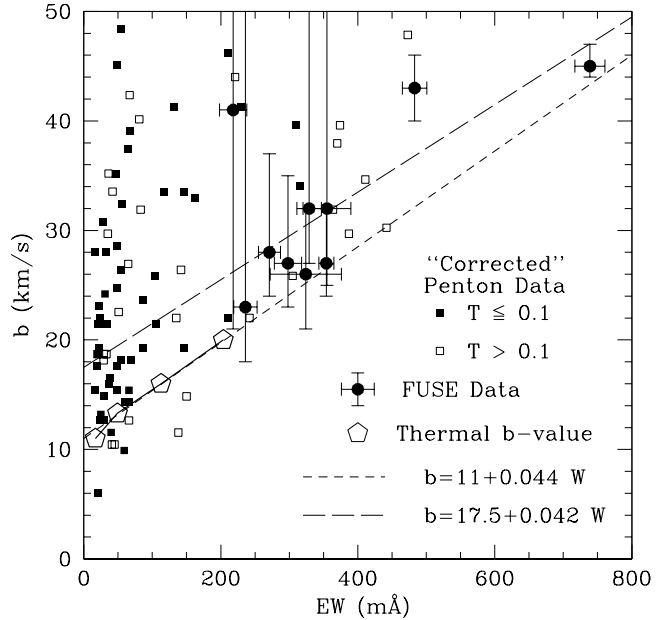}
\caption{ \label{fig-bscatt} The $b$-value data of \citet{Shull:00}
using \emph{FUSE} (large solid circles with error bars), plotted
against the rest EW.  These results imply that intrinsic $b$-values
for clouds with ${\mathcal W} \ge 200$ m\AA\ are roughly half of the
observed $b_{Ly \alpha}$ (see \S\ref{sec-EW-colden}).  The cloud data
are presented as small squares; larger tidal fields (${\mathcal T} \ge
0.1$) are open squares, and for ${\mathcal T} \le 0.1$, solid squares.
These Doppler parameters have been reduced to 55\% of their $b_{Ly
\alpha}$ value, according to this prescription.  Open pentagons
represent a theoretical lower limit to the observed $b$-value, based
on the column density-weighted temperature of the model clouds for
log-columns of 12.5, 13.0, 13.5, and 14.0, left to right (see
\S\ref{sec-EW-colden}).  The short-dashed line is an extension of the
trend of the pentagons, showing that the FUSE data obeys the envelope.
The long-dashed line represents the approximate median relationship
(see text).  See text for details. }
\end{figure}

In Fig. \ref{fig-bscatt}, the derived $b$-values of clouds from the
\emph{FUSE} analysis are displayed as filled circles together with
their 1-$\sigma$ errors in ${\mathcal W}$ and $b$.  The lower limit of
their distribution should be limited by the thermal equilibrium, as
the turbulent broadening approaches zero.  A theoretical limit for the
thermal broadening for lower EW clouds can also be applied by noting
the column-weighted temperatures of model absorption systems, as
discussed in \S\ref{sec-T-b}, and plotted as open pentagons in the
figure.  The apparent linear trend of the latter three points is
extended (dashed line) to high EW.  It is seen that the derived
$b$-values of large EW clouds obey the lower-limit that thermal
broadening in absence of turbulence would imply. The equation of the
line is,
\begin{equation} \label{eq-b-vs-W1}
b=11+0.04375 \,{\mathcal W} ~\kms.
\end{equation}

Also shown in Fig. \ref{fig-bscatt} are the cloud data of
\citet{Penton:00a}, which have been reduced to 55\% of their recorded
$b_{Ly \alpha}$ values, in anticipation of the applicability of a
similar correction factor for lower EW clouds (see \S\ref{sec-bval}).
The appearance of a few clouds below the line of thermal broadening
limit should not alarm the reader; in applying the mean reduction
factor, we are introducing an error of order 30\%; the fractional
dispersion in the relation of $b$ to $b_{Ly \alpha}$ from the
\citet{Shull:00} data.  For a cloud of true $b$-value 15 \kms, the
1-$\sigma$ error is of order 4.5 \kms, enough to displace a cloud with
small bulk motions to well-below the line.

Figure \ref{fig-bscatt} demonstrates that $b$-values of clouds can be
characterized as having a clear minimum for a given EW.  Clearly,
though, this will over-estimate the column density of an average
cloud.  However, a median b-value line may be drawn (long-dashed line)
as a function of EW that is more relevant for transforming the EWDF
to the CDS since it approximates the most probable $b$-values.  This
line has the equation
\begin{equation} \label{eq-b-vs-W2}
b\approx17.5+0.042 \,{\mathcal W} ~\kms,
\end{equation}
and is designed to represent the contributions of thermal broadening
as well as bulk motions or the effects of unresolved components.

\begin{figure}[h!] 
\centering
\epsscale{1.0} 
\plotone{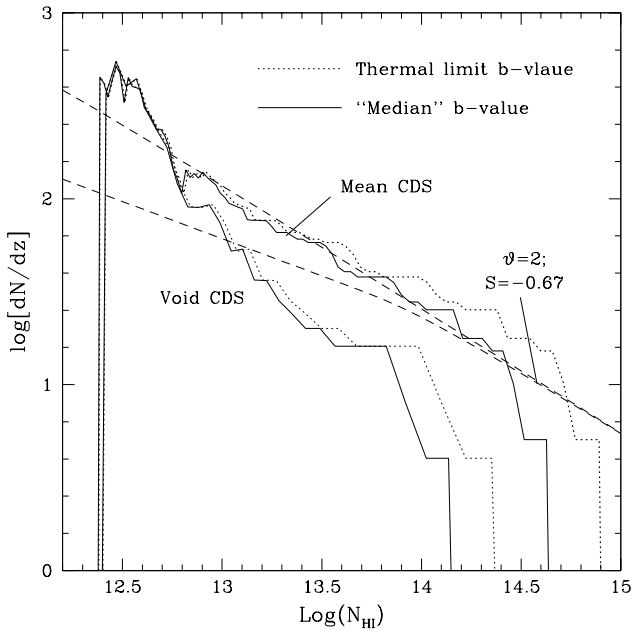}
\caption{\label{fig-CDS-varb}The observed column density spectrum
derived using Eq. \ref{eq-b-vs-W1} (dotted lines), and
Eq. \ref{eq-b-vs-W2} (solid lines), for the volume-weighted column
density distribution (upper pair of lines) and the void CDS (${\mathcal
T} \le 0.03$, lower pair).  These distributions are compared to an
analytic fit of arbitrary normalization for an untruncated isothermal
baryon distribution (upper dashed line) and a truncated isothermal
model with $r_t\simeq 144 ~\kpc$ for a halo with $v_c \simeq 21$ \kms\
(lower dashed line).  The effect of truncation is
to diminish the number of low column density absorbers.  The slope of
the untruncated line is ${\mathcal S}=-0.67$. }
\end{figure}

The result of applying these formulaic relations to the clouds via
Eq. \ref{eq-cherno} results in the mean, and void EWDFs
(Fig. \ref{fig-ewds2}) being transformed into a CDS.  This is shown in
Fig. \ref{fig-CDS-varb}, where the dotted lines represent the results
of applying Eq. \ref{eq-b-vs-W1} to void, and mean cloud catalogs, and
the solid lines represents the application of Eq. \ref{eq-b-vs-W2}.
The dashed lines accompanying these plots represent the CDS that would
result from an untruncated isothermal sphere ($\vartheta=2$, resulting
in a spectral slope ${\mathcal S}\simeq-0.67$ top line), and a
truncated isothermal sphere (lower dashed line), where the truncation
radius scales with $r_{t}\simeq 144 ~\h_{75}^{-1}$ \kpc\ for a cloud
circular velocity $v_c=21 $ \kms, about twice the scaled truncation
radius assumed in Paper 1 (where $R_t = 500 X_v h_{75}^{-1} ~\kpc$).
At this distance the overdensity of baryons and DM (assuming baryons
follow DM) is $2.57 ~\Omega_b ~\rho_{crit}$, and thus shows the strong
effect of a truncation (such as from ram pressure stripping of
diffuse, dissipated gas in filaments) on the CDS slope at low
$N_{HI}$.

For the purposes of transforming an EWDF into a CDS, or vice versa the
median relationship of $b({\mathcal W})$ (Eq. \ref{eq-b-vs-W2}) will
henceforth be used.  Though the error at column densities $N_{HI} \ga
10^{13.5} $ \cmtw\ may be substantial, smaller columns will be
minimally affected, as can be seen in Fig. \ref{fig-CDS-varb}.

\section{Modeling distributions of subgalactic halos} \label{sec-modeling}

The foregoing developments now enable the modeling of the CDS with
distributions of model clouds.  Cloud profiles can be combined into a
column density spectrum, given a distribution of halo velocities.  The
primary method of determining the halo distribution function used here
involves transforming the luminosity function (LF) into a halo
velocity distribution function (hereafter HVDF).  Additional useful
information can be garnered by comparing our results to those emerging
from numerical simulations.

The required procedure can be outlined as follows: First, extract the
column density of a model cloud as a function of projected radius from
a $z=0$ product of the 1-D simulation.  Secondly, combine this with a
distribution of halos, and compare this synthetic spectrum to
observations.

Since clouds are here assumed to be sub-galactic perturbations, I
approach the subject by transforming the LF of galaxies into a HVDF
that can be used to synthesize a CDS out of model clouds.  The halo
mass and velocity distribution functions of galaxies are determined
by applying the Tully-Fisher (TF) relation to the galaxy luminosity
function.

The TF relation be used to transform the luminosity function into an
HVDF.  It can be represented in many ways.  Perhaps the simplest is as
the power-law,
\begin{equation} \label{eq-tf}
\frac{\mathcal L}{{\mathcal L}^*} =
\biggl(\frac{v_c}{v_c^*} \biggr)^{\beta},
\end{equation}
where $v_c^*=161 \,$ \kms\ (see \S5.1 of Paper 1).  The value of the
constant $\beta$ is a function of the photometric band being used.
For the $B$-band, $\beta=2.91$, and for the $K$-band, $\beta = 3.51$
\citep{Tully:00}.  Note that all symbols with a star, \eg, $\,M^*$,
$R_t^*$, \& etc., refer to expectation values for an ${\mathcal L}^*$
system, itself referring to the point of inflection of the Schechter
form of the LF (\ie, the ``knee'').

As long as there is a calibrated TF relation corresponding to (\ie, in
the same spectral bandpass as) a given LF, one may apply a
transformation of the LF into a mass, or velocity-DF.  Below, is
detailed the application of this transformation.

\subsection{Transforming the LF to a Velocity or Mass DF} \label{sec-trans-LF}
The halo velocity or mass distributions are based on the luminosity
function.  For these distribution functions I make the following
syntactical simplifications,
\begin{eqnarray} \label{eq-Xdefs}
X_{\mathcal L} &=& \frac{\mathcal L}{{\mathcal L}^*}, \nonumber\\
X_{v} &=& \frac{v_c}{v_c^*},\\
X_{\mathcal M} &=& \frac{{\mathcal M}}{{\mathcal M}^*} \nonumber,
\end{eqnarray}
where ${\mathcal M}$ represents the halo mass.
The Schechter luminosity function is,
\begin{equation} \label{eq-SLF}
\phi_{\mathcal L} = \phi_{\mathcal L}^*\, X_{\mathcal L}^{\alpha}\, e^{- X_{\mathcal L}},
\end{equation}
\citep{Schechter:76} and is a differential DF.  A 1-D analogue to the
Jacobian will transform the LF into something else.  For instance, to
transform a Schechter distribution $\phi_{\mathcal L}$ into a HVDF
($\phi_{v}$) we have, $\phi_{v}=\phi_{\mathcal L}\partial{X_{\mathcal
L}}/\partial{X_{v_c}}$.  Partial derivatives are calculated from the
following scaling relations for a truncated ISOT halo (see Paper 1, Eq. 29),
\begin{equation} \label{eq-scale-ISO}
X_v = \frac{R_t}{R_t^*} = X_{\mathcal L}^{1/\beta} = X_{\mathcal M}^{\frac{1}{3}}, 
\end{equation}
where $R_t$ is the galaxy halo truncation radius.  The truncation
radius utilized here, for clouds as well as galaxies, is such that the
total mass density at $R_t$ is $\rho(R_t)=10 ~\rho_{crit}$.  For the
NFW halo the following scaling relations obtain (Paper 1, Eq. 40),
\begin{equation} \label{eq-scale-NFW}
X_v =\biggl(\frac{R_{max}}{R_{max}^*}
\biggr)^{0.67} =\biggl(\frac{R_{vir}}{R_{vir}^*} \biggr)^{0.76} =
X_{\mathcal L}^{1/\beta} = X_{\mathcal M}^{0.30}.
\end{equation}
As with Paper 1, we correct both the ISOT and NFW halos for the amount
of baryons expected to be associated with it according to the assumed
value of $\Omega_b/\Omega_m$.  With baryons included, $v_c^*=161.5
$ \kms\ (see \S5.1.1 of Paper 1), but the dark halo has a halo
$v_c^*=144 $ \kms.  

Since $X_{\mathcal L} =X_v^{\beta}$ (Eq. \ref{eq-tf}),
\begin{eqnarray} \label{eq-HVDF}
\phi_v = \phi_{\mathcal L} \frac{\partial X_{\mathcal L}}{\partial X_v} = 
\beta\phi_{\mathcal L}^* X_v^{\beta(\alpha+1)-1} e^{-X_v^{\beta}}.
\end{eqnarray}
This halo velocity distribution function is identical for ISOT and NFW
halos.  The mass function for the ISOT halo is,
\begin{equation} \label{eq-massfunc-ISO}
\phi_{\mathcal M} = \phi_{\mathcal L} \frac{\partial X_{\mathcal L}}{\partial X_{\mathcal
M}} = \left(\frac{\beta \phi_{\mathcal L}^*}{3} \right) X_{\mathcal
M}^{\beta(\alpha+1)/3 -1} e^{-X_{\mathcal M}^{\beta/3}},
\end{equation}
while for the NFW halo we have,
\begin{equation} \label{eq-massfunc-NFW}
\phi_{\mathcal M} = \phi_{\mathcal L} \frac{\partial X_{\mathcal L}}{\partial X_{\mathcal
M}} = \left(0.3 \beta \phi_{\mathcal L}^* \right) X_{\mathcal
M}^{0.3 \beta(\alpha+1) -1} e^{-X_{\mathcal M}^{0.3 \beta}},
\end{equation}
Note that the exponential may be safely ignored for clouds, that have
$X_v$ and $X_{\mathcal M} \ll 1$.

The normalizations for each of these distributions can easily be
calculated using the normalizations of the luminosity function.  For
the B-band LF, it is thought wise to interpolate the bracketing Sloan
Digital Sky Survey (SDSS) normalizations to the B-band owing to the
apparent consistency of their normalizations (see Appendix of Paper 1)
and the large database used.  We therefore use, 
\begin{equation} \label{eq-norm}
\phi_B^*=0.022 \, {\rm h}^3 \,{\rm Mpc}^{-3}= 0.0093 \, {\rm h}_{75}^3
\,{\rm Mpc}^{-3}.
\end{equation}

The distribution functions given in this section provide the basic
framework for placing our halos into the universe.  Although there are
some conflicting indicators of which value of the slope $\alpha$ to
apply, we may either prevail upon the criterion that $\alpha$ should
be steep enough to explain the void cloud line density, or find a
preferred value for $\alpha$ based on other compelling work.

\subsection{Comparison with Previous HVDF Simulations} \label{subsec-comp-Klypin}
Cosmological simulations of hierarchical clustering
\citep[\eg,][]{Klypin:99, Klypin:02} (hereafter K99, and K02,
respectively) provide the basis for deriving velocity and mass
distribution functions of halos.  How do the slopes of these
distributions relate to the functional forms derived above?  K02 noted
an environmental dependency in the halo velocity distribution, with
slopes $-3.1$ in groups, and $-4.0$ for isolated halos.  With these
facts we can find the Schechter faint-end slope parameter $\alpha$
that is consistent with these values.  For small halo velocities, the
HVDF (Eq. \ref{eq-HVDF}) can be simplified by removing the exponential
which is $\sim 1$ for cloud halos.  The HVDF is the same for ISOT and
NFW, but there are two different environments; the dense (\ie, group),
and the isolated environments.  I alternately equate the HVDF slope,
$\beta(\alpha+1) - 1$, to $-3.1$, and $-4.0$, and derive the slope
parameter using $\beta_B=2.91$: $\alpha = -1.72$, and $-2.03$, for
group, and isolated halo distributions, respectively.  Note that these
simulations presume NFW halos.  It stands to reason that the isolated
halo distribution would be more applicable to void cloud
distributions.

By equating the mass function slopes in Eqs. \ref{eq-massfunc-ISO} and
\ref{eq-massfunc-NFW} (excised of their exponentials) to that of the
differential mass function with a slope of $-1.8$, as found by K02, I
find, $\alpha=-1.83$ and $-1.93$, for ISOT, and NFW, respectively
($\beta_B=2.91$).  These represent values a mixture of isolated
and group halos.

If isolated NFW halos have a slope parameter $\alpha=-2.03$, perhaps
the slope slope parameter for isolated ISOT halos can be estimated by
the difference between ISOT and NFW slopes for mixed environments
($\alpha=-1.83$ and $-1.93$, respectively, as above).  Since the ISOT
slope was flatter than the NFW mass function by 0.1 for mixed
environments, then this would imply that for isolated ISOT halos,
$\alpha=-2.03+0.1 \simeq -1.93$.  I consider a slightly rounded value
$\alpha= -1.95$ to be a provisional order-of-magnitude prediction for
ISOT clouds in voids.

This steep slope, in comparison to a faint-end slope of $\alpha
\approx -1.25$ for galaxies (\citealp{Blanton:01}), demonstrates the
source of the ``missing galactic satellites'' problem (\eg, K99).  The
suggestion of K99 is that the missing satellites may be clouds, rather
than dwarf galaxies.  In this case, their work may imply that the
parameter $\alpha$, if it is taken as referring to the demographics of
halos in general, rather than the luminosity of halos, may be
indicative of the presence of numerous small halos that are not
luminous.  Whether these are LSB galaxies, HVCs or \lya\ clouds, a
steep EWDF slope parameter suggests that a significant amount of mass
may be tied up either in small halos, or in an unconsolidated
component, as we shall see in \S \ref{sec-mass-dist}.


\subsection{Assembling the CDS} \label{sec-CDS}
The column density spectrum is calculated from model clouds according
to the equation,
\begin{equation}
\frac{d {\mathcal N}(\ge N_{HI})}{d \, z} =
\frac{c}{H_0} \sum_{n=0}^{13} \phi(X_v(n)) ~\pi
\,r_{N_{HI}}^2 \delta{X_v(n)},
\end{equation}
where $\phi(v_c)$ is given by Eq. \ref{eq-HVDF}, $r_{N_{HI}}$ is the
radius at which a column density $N_{HI}$ is detected, ${X_v(n)}$ is
given by Eq. \ref{eq-bin-cntr2}, and $\delta {X_v(n)}$ is the bin
width.  The cadence in $X_v$ indicates that $\delta{X_v} = X_v
(10^{0.025}-10^{-0.025}) = 0.1152 \, X_v$.  The radius at which a
given column density can be observed in a cloud of a given $v_c$ is
determined by the method described in \S\ref{sec-rp-colden2}.
The model column density spectra can then be compared with the
observations.


The goal of modeling the distribution of clouds is to reproduce the
observations.  Not only must the line densities as a function of
column density be matched, but it is also important to reproduce the
shape of the CDS.

Equation \ref{eq-HVDF} shows how the halo distribution function is
calculated as a function of $X_v$, and demonstrates the role of the
faint end slope $\alpha$ and the TF slope $\beta$ in determining the
low-velocity end of the HVDF.  Since it is theorized that the
distribution of sub-galactic halos is in continuity with the galaxy
halo EWDF, the normalization of the galaxy luminosity function in
voids will be a good place to start the modeling; the modeled
distribution of halos must be based on a normalization $\phi^*(void)$.
It has been suggested that galaxies are present in voids at $\sim
1/10$ the average density \citep{El-Ad:97b, El-Ad:00}.  A somewhat
larger galaxy density is found in $\rm{ Bo{\ddot{o}}tes}$
\citep{Dey:95, Cruzen:02}.  An upper-limit of current thinking might
be represented by the simulations of \citet{Cen:99}, which assumes
that $\Omega_m=0.4$ and estimates that $\Omega_V/\Omega_m \simeq
0.26$, consistent, perhaps, with roughly an equal mass in clouds and
in galaxies.  A low normalization would seem to imply a low total void
density, yet since the spectral slope of the void cloud HVDF may be
steep, this is not necessarily inconsistent with a large density.
Therefore, I provisionally adopt the normalization, $\phi^*(void) =
0.1 \phi^* = 9.3 \times 10^{-4} \, {\rm h_{75}^{-3} \, Mpc^3}$ (see
Eq. \ref{eq-norm}), with hopes that it can explain the data.

The somewhat arduous transformation of the TF relation to an HVDF, and
the number of parameters that need to be evaluated, may give the
impression that there are many free parameters utilized in this
analysis.  This is a misconception, for none of the parameters are
given arbitrary values.  The slope parameter $\alpha$ is reasonably
constrained by the modified Press-Schechter simulations of K02 (see
\S\ref{subsec-comp-Klypin}), the TF slope $\beta$ is constrained by
the B-band TF relation, which has a well-known value \citep{Tully:00},
and the normalization of the void EWDF is observationally constrained,
as noted above.  Of these, the latter would seem to have the greatest
range of uncertainty, and is the only parameter used as though it were
a truly free parameter.  However, none of the values of these
parameters are arbitrary, or disconnected from observations.

The effects of using various slope parameters $\alpha$ on the derived
CDS are discussed in the following section.

\subsection{The model CDS with slope parameter $\alpha$}
A constant slope parameter $\alpha$ is the most obvious beginning
point for investigating the relationships between our parameters and
the product CDS.

\subsubsection{The ISOT model}\label{sec-ISOT}
\begin{figure}[h!] 
\centering
\epsscale{1.0} 
\plotone{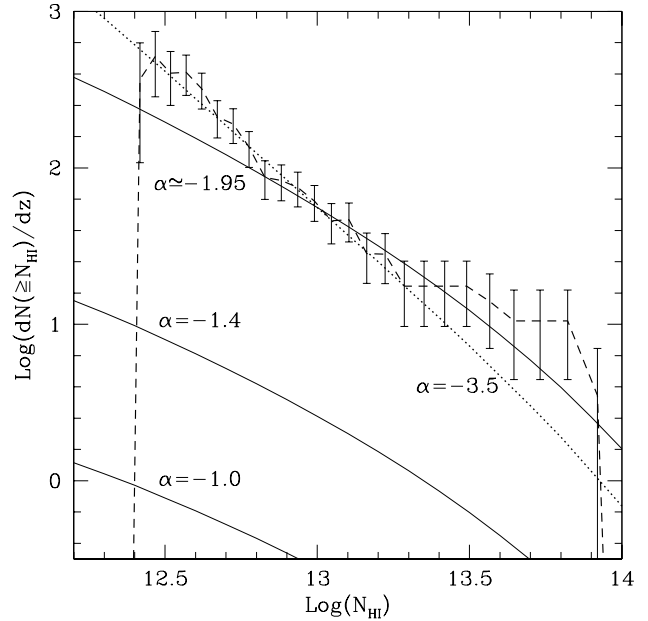}
\caption{\label{fig-CDS-const-alph} A comparison of the observed void
column density spectrum for low-$z$ clouds (${\mathcal T} \le 0.1$)
(short-dashed line) with spectra that would result from a HVDF
resulting from an extension of the SLF faint--end slope $\alpha=-1.0$,
$-1.4$, and $-1.95$, normalized at $0.1 \phi^*$ (solid lines).  To fit
the slope at $N_{HI} \la 10^{13.3} $ \cmtw, an HVDF with a slope
parameter $\alpha \simeq -3.5$ is required (dotted line), though the
required normalization is an order of magnitude lower than $0.1
\phi^*$.}
\end{figure}

I first consider the ISOT model.  Recall that with the ISOT halo,
clouds with large $v_c$ significantly restrain the expansion of
baryons from the dark halo (\S\ref{sec-ISOT-profiles}).  As a result,
columns $N_{HI} \ga 10^{12.4}$ \cmtw\ are detected for the velocity
range $10 \le v_c \le 25$ \kms.  Figure \ref{fig-CDS-const-alph}
displays model CDS (solid curves) for various slope parameters
$\alpha$ using the above normalization and the ISOT model profiles
(Fig. \ref{fig-deniso}).  When applying slopes of $\alpha=-1.0$ or
$-1.4$, low line densities result.  If the slope is $-1.4$, then the
normalization will have to be increased by a factor of 30 in order to
give rough agreement with the observed line density, making the
normalization in void space greater than the average by a factor of 3,
which is absurd.  When $\alpha \approx -1.95$, as suggested in
\S\ref{subsec-comp-Klypin}, the resulting modeled line density is
comparable to the observed value.  In order to fit the low column
density end of the CDS, a slope parameter of order $\alpha \approx
-3.5$ is required (dotted line in Fig. \ref{fig-CDS-const-alph}).
However, integrating over the HVDF (for details, see
\S\ref{sec-mass-dist}), the total mass density for clouds $5 \le v_c
\le 25$ \kms\ is $\Omega_{cl} \ga 0.028$, of order 10\% of the assumed
mean density $\Omega_{m}=0.3$, when the truncation radius is taken to
be $r_t = 500 X_v ~\h_{75}^{-1} ~\kpc$, where $X_v$ is given by
Eq. \ref{eq-scale-ISO}, as was the standard with Paper 1 (\S5.1.1).
When systems of velocity $10 \le v_c \le 150$ \kms\ are used, then
$\Omega_V=0.09$, and $\Omega_V/\Omega_m \simeq 0.3$.  The truncation
radius is such that the density at $r_t$ is about 10 times the
critical.  Considering that the rest of void space is unlikely to be
absolutely empty, this value for $\Omega_V$ must be a lower limit.

\subsubsection{The NFW model}
NFW halos do not produce any absorbers with $N_{HI} \ge 10^{12.4} $
\cmtw\ in the range $10 \le v_c \le 22.4$ \kms; the next higher
velocity bin ($n=13$ in Eq. 20) experiences collapse in the center,
presumably forming stars.  The line density, even at $N_{HI}=10^{12.2}
$ \cmtw\ is less than one cloud per unit redshift for parameters that
produce an acceptable fit with ISOT halos.  The reasons for this
shortfall are two-fold.  First, NFW halos are about 5 times less
massive than an ISOT halo of the same $v_c$, and hence, have 5 times
fewer baryons.  Second, the lower mass of the halos cannot as
effectively slow the baryonic outflow and dispersal as can the ISOT
model; the trend of the restraint of the outflowing gas for large
$v_c$, noted in the inset of Fig. \ref{fig-deniso}, is not apparent in
the NFW profiles (Fig. \ref{fig-den-NFW}).  Further, comparison of
Figs \ref{fig-deniso} and \ref{fig-den-NFW} shows that the central
baryonic densities of NFW halos are of order 1/10 of an ISOT halo at
comparable halo velocities.  Since no absorption systems are produced,
this halo model fails to explain the observed EWDF, and I forego any
further consideration of it.

\subsection{The Shape of the CDS} \label{sec-shape-CDS}

In this section I take a critical look at the observed CDS to see what
degree of confidence is justified in its slope at the low, or ``high''
column density end ($\sim 10^{14} ~\cmtw$).  Figure
\ref{fig-CDS-const-alph} shows the Poisson error bars for data at $z
\leq 0.036$ (see \S6.2.1 of Paper 1).  Adjacent error bars are not
independent of each other since the same clouds may contribute to a
range of EWs.  When data from all redshifts are used, the Poisson
errors are smaller, but the slope at the low column-density end
(${\mathcal W} \la 40$ m\AA, or $N_{HI} \la 10^{13.3} $ \cmtw) is
moderately flatter than if only data from $z \le 0.036$ is used (see
Fig. 11 and 12 of Paper 1).  This is probably attributable to an
underestimation of the tidal field at higher redshifts, with
consequent contamination of void catalogs with non-void clouds, all
occasioned by the probable non-detection of less-luminous galaxies in
the proximity of the clouds.  I conclude that the steepness of the
slope apparent at $\log(N_{HI}) \la 13.3$ in
Fig. \ref{fig-CDS-const-alph} is highly significant, though what it
means is yet to be determined (see \S\ref{sec-mass-dist}).

However, the ``bump'' in the CDS at columns $N_{HI} \ga 10^{13.5} $
\cmtw, remains essentially unchanged when higher redshift data are
excluded.  Is this a real feature?  There are five clouds with $N_{HI}
\ge 100$ m\AA\ (which, using Eqs. \ref{eq-b-vs-W2} and
\ref{eq-cherno}, implies $N_{HI} \ga 10^{13.35} $ \cmtw) contributing
to this end of the low-$z$ void distribution (${\mathcal T} \le 0.1$),
as opposed to $n_{cl} \sim 20$ at $N_{HI} \la 10^{12.9} $ \cmtw.  It
is helpful, at this point, to take a detailed look at the five clouds
with ${\mathcal W} \ga 100$ m\AA\ in the low-$z$ data with the help of
a program that assesses the prospects for the physical association of
clouds (\ie, gravitationally bound) with galaxies on the basis of the
galaxy absolute magnitude (mass is derived assuming the more massive
ISOT halo), the relative velocity of cloud and galaxy, and the
projected radius, (Manning, in preparation).  The galaxies considered
for this routine are those within 750 \kms\ of the cloud in question
if the projected radius $r_p \le 2$ Mpc, or within 350 \kms\ if
outside this.

Two of the five low-$z$ clouds have derived local tidal fields of
${\mathcal T}=0.073$ and 0.098.  Both are found to have one
sub-${\mathcal L}^*$ galaxy at a projected distance $r_p > 2$ Mpc,
with cloud-galaxy LOS relative velocities of 66 and 72 \kms,
respectively, for which the program predicts probabilities of physical
association of $\sim 13\%$, and 4\%, respectively.  These clouds can
be found in Fig. 4 of Paper 1 in the sightlines to MRK 335 (Fig. 4k,
1241.09 \AA) and AKN 120 (Fig. 4h, 1247.94 \AA), respectively.  But
even this possible physical association could not produce a
significant peculiar velocity along the LOS because of the large
projected distance to the galaxy, and hence the attributed tide cannot
be far wrong.  Of the other three, two have tidal fields ${\mathcal T}
\approx 1 \times 10^{-3}$, and have no galaxies within the search
area.  The other, with ${\mathcal T} \simeq 7\times 10^{-3}$, has a
late-type sub-${\mathcal L}^*$ galaxy at a projected radius $r_p
\simeq 430$ \kpc\ (the CfA Redshift Catalog, \citealt{Huchra:90},
\citealt*{Huchra:95}, \citealt*{Marzke:96}, \citealt*{Grogin:98},
\citealt*{Huchra:99})\footnote{see
http://cfa-www.harvard.edu/~huchra/}, but at a relative velocity of
$\Delta v \simeq 350$ \kms.  The probability of association was rated
at $\sim 4\%$. The galaxy velocity error was listed as $\delta v = 1$
\kms, and the cloud redshift error from \citet{Penton:00a} was listed
as $\delta v=6$ \kms.  Thus, these five clouds have few prospects for
a physical association in which significant peculiar velocities from
gravitational interactions could result, and fewer chances for
associations that would give significant LOS peculiar velocities.
With high confidence, I claim these 5 clouds are truly remote, and are
not anomalously present in void catalogs.

It is noteworthy that the void filling factor at ${\mathcal T} \la
0.01$ is $\approx0.44$ (see Fig. 7, Paper 1), while at ${\mathcal T}
\la 0.1$, $f_V\simeq 0.86$.  Thus the volume in which the three low
${\mathcal T}$ clouds appear is almost exactly the same as the volume
in which the two at $0.01 \le {\mathcal T} \le 0.1$ clouds appear,
making it seem likely that large clouds may be more prevalent in deep
voids than in the void periphery.  But the statistics are small, and
we would be wise to seek more data before forming a conclusion.

A possible explanation for those clouds that produce the excess above
the model at $N_{HI} \approx 10^{13.7}$ (see
Fig. \ref{fig-CDS-const-alph}) is that there exist
clouds/proto-galaxies in voids with halo velocities $v_c > 25$ \kms,
for the analysis of the CDS of individual clouds shows that the larger
the circular velocity of a cloud, the flatter the slope of the CDS
that it produces; such additions would produce more absorbers at $\log
{N_{HI}} \sim 13$ to 14 (\cmtw).  This can be understood by noting
that the more massive the cloud, the more strongly it will hold its
baryons, producing a steep density gradient at the edge, and resulting
in a flat CDS (\cf Eqs. \ref{eq-CDS-slope1} and \ref{eq-CDS-slope2}).

Recall that the gas in larger halos ($v_c \ge 25$ \kms) collapse and
produce stars in the 1-D simulations.  However, if clouds have grown
from hierarchical merging during their evolution, they would have been
smaller during the early, high-density stages of evolution, achieving
their larger size only recently.  Because of the early evaporation of
baryons from their core region, there might not be sufficient baryonic
mass in the core to produce the cooling needed to trigger star
formation.  Therefore, with growing halos, the limiting circular
velocity (as measured at $z=0$) for SF is probably higher than 25
\kms.

But before pursuing this line of thought, it is important to consider the
repercussions of taking literally the significant steep CDS slope at
$N_{HI} \la 10^{13.3} $ \cmtw\ noted above.

\subsection{Total mass distribution and slope parameter $\alpha$} \label{sec-mass-dist}

As noted in \S\ref{sec-ISOT}, the low column density end of the void
CDS can be well-fit by an HVDF with a very steep slope parameter
$\alpha \approx -3.5$.  In this section I inspect the physical
repercussions of this.  As argued in \S\ref{subsec-comp-Klypin}, the
HVDF may be simplified by removing the exponential (Eq. \ref{eq-SLF})
when $X_v \ll 1$.  Consider logarithmic bins of $v_c$ such as
was used in the simulations here; construct the bins such that each
bin has a width $\delta\,X_v\equiv X_v \,\delta v_0/v_0$, where
$\delta v_0/v_0 \ll 1$ is some appropriate constant.  Then from
Eq. \ref{eq-HVDF}, the number density of systems within a bin is,
\begin{equation}
\delta n_{bin} = \phi_v^* X_v^{\{\beta(\alpha+1)-1\}} X_v
\left\{\frac{\delta v_0}{v_0}\right\}.
\end{equation}
Thus, the comoving number density per bin scales as
\begin{equation}
\delta n_{bin} \propto X_v^{\beta(\alpha+1)}.
\end{equation}
When $\alpha=-1$, each logarithmic bin has an equal number
of systems in it.  However, when we consider the mass in each bin, we
must multiply the number density by the mass per object.  

For the ISOT halo, the mass scales as $M \propto X_v^3$ (Eq. \ref{eq-scale-ISO}),
\begin{equation}
\delta M_{bin} \propto X_v^{\beta(\alpha+1)+3} \propto X_{\mathcal
M}^{\beta(\alpha+1)/3+1},
\end{equation}
that has equal masses in logarithmic bins when $\alpha =
-(1+3/\beta)\simeq -2.03$ for $\beta_B=2.91$.  For this value of
$\alpha$, the mass function is flat in logarithmic bins.

For the NFW halo, virial mass $M_{200} \propto v_{max}^{3.30}$
Eq. \ref{eq-scale-NFW}).  The NFW bin mass is a somewhat different relation,
\begin{equation}
\delta M_{bin} \propto X_v^{(\beta)(\alpha+1)+3.3} ,
\end{equation}
that has equal masses in logarithmic bins when $\alpha=-3.3/\beta -1
\simeq -2.13$ for the same $\beta_B$.  

Recall that a CDS formed using ISOT cloud profiles and a HVDF
with $\alpha=-3.5$ would have an increasing mass in bins of
progressively smaller velocity $v_c$, implying a divergence of the
integrated mass.  The sobering implications of this fact are softened,
however, if sub-clustering is occurring.  Further, it is reasonable
that sub-halos should form in the flanks of larger halos, for the
time-scale for collapse of perturbations is proportional to the
inverse square root of the density, making it more likely for
them to form there than in the regions between clouds.  Thus, many of
the small clouds that are detected in the CDS may not be independent,
isolated clouds, but loosely associated with larger clouds.  Since
they are fashioned out of the stuff of the main cloud, they are not
taken to add to its mass, unless such sub-halos are found to lie
outside the truncation radius $R_t$.  If sub-condensations exist, then
accretion or hierarchical agglomeration is a distinct possibility in
the halos of \hone\ clouds in voids.

\section{Extending Model Clouds to Higher $v_c$} \label{sec-extending}

Up to this point, the halos that have been used were fixed at $z=6.5$.
Such halos remained star-free only up to circular velocities of only
up to $v_c \approx 25~$ \kms\ (\S\ref{sec-results}).  However, the
steep spectral slope of the void CDS at low column density appeared to
require the presence of sub-clustered halos, since otherwise the
indicative mass function would diverge.  Likewise, it is difficult to
reconcile the existence of CHVCs \citep{Braun:00} with \hone\
mass approaching $10^7 ~\Msun$ \citep{deHeij:02}, with model clouds
having maximum \hone\ masses only $\la 10^6 ~\Msun$.

\begin{figure}[h!] 
\centering
\epsscale{1.0} 
\plotone{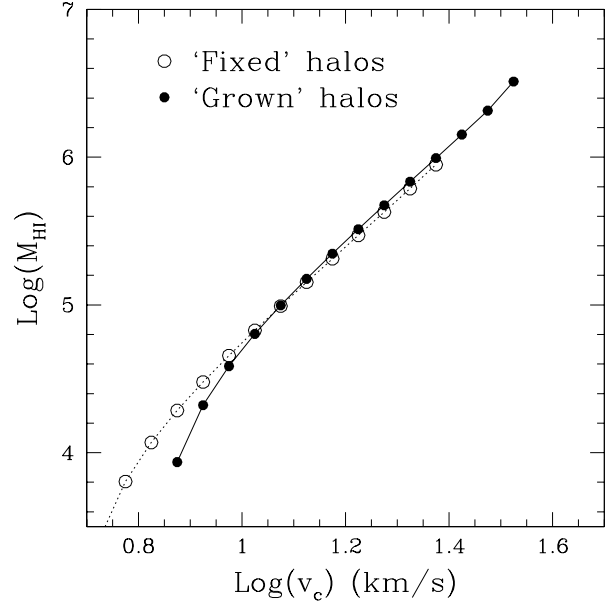}
\caption{\label{fig-mh1} The \hone\ mass in terms of the circular
velocity for clouds that grow according to Eq. \ref{eq-grow} (filled
circles and solid line) compared to that of the fixed halo (open
circles, dotted line).  The logarithm of the \hone\ mass is shown as a
function of the logarithm of the cloud circular velocity (solid line) for
untruncated clouds.
}
\end{figure}

The behavior of the observed CDS relative to model CDS thus suggests
the possibility of continued accretion and halo growth during the
period from $z=6.5$ to the present; that sub-halos cluster about
larger halos implies that the circular velocities of clouds could grow
by the accretion of those halos; their growth would explain the
presence of CHVCs\footnote{the optical surface brightness of
local CHVCs is constrained to be $\la 10\%$ of the faintest
Local Group dSph (L. Blitz, private communication 2002)} with \hone\
masses approaching $10^7$ \Msun.  Figure \ref{fig-mh1}, shows \hone\
mass for fixed halos (open circles), corresponding to each circular
velocity bin).  The trend of fixed halos shown in
Fig. \ref{fig-mh1}, halos of velocity $v_c \ga 40$ \kms\ could reach
this \hone\ mass.

Recall, however, that attempts to run models of such large halo
velocities failed for fixed halos, because the central baryonic core,
thought to roughly follow the dark matter before reionization, was not
able to support itself, as described in \S\ref{sec-results}.  However,
if halos are slowly ``grown'', such as by accretion, then larger halo
velocities at $z=0$ may be attainable without these indications of
star formation, with \hone\ masses more in line with observations of
CHVCs may be attained.

In what follows, I adopt this model of accretion.  A rough
approximation of that growth rate was made by assuming that the largest
fixed cloud that escapes collapse into star formation, could grow
to the largest cloud which today has an \hone\ mass of
$10^7 ~\Msun$, as asserted by \citet{Braun:02}.  The circular velocity
of that halo was approximated by connecting the trend in \hone\ mass
with fixed-halo circular velocity to the maximum cloud \hone\ mass of
$10^7 ~\Msun$ (see Fig. \ref{fig-mh1}).

\begin{figure}[h!] 
\centering
\epsscale{1.0} 
\plotone{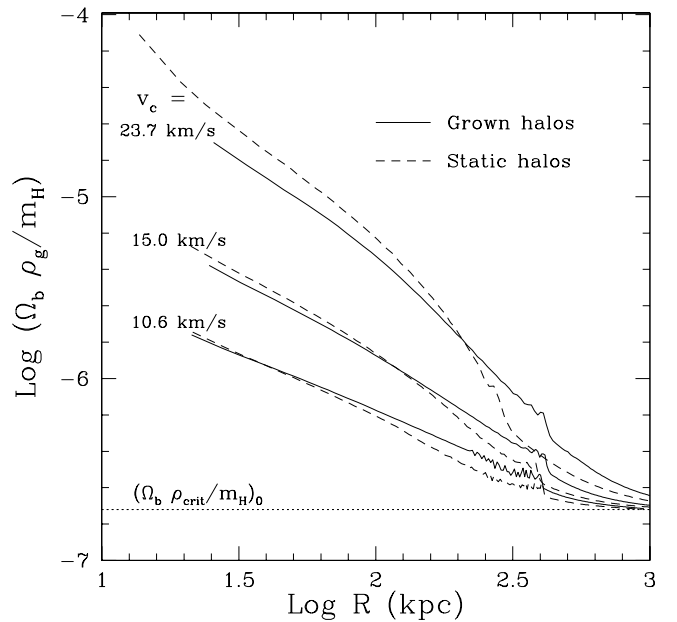}
\caption{\label{fig-g-ng} Density profiles for clouds with circular
velocity $v_c \simeq 9.4$, 15.0, and $23.7 $ \kms\ for  fixed halos
(dashed lines) and for halos which grow according to Eq. \ref{eq-grow}
(solid lines).  The more power-law like profile results in a steeper
CDS slope (see Fig. \ref{fig-CDS-g-ng}).  }
\end{figure}

The clouds are ``grown'' gradually, that is, by inserting increments
of mass into the Lagrangian bins at each time step in a way that
mimics the equation,
\begin{equation} \label{eq-grow}
v_c(t)=v_c(0) \, e^{k_a t},
\end{equation}
where $v_c(0)$ is the circular velocity at $z=6.5$, the accretion
constant $k_a \approx 0.78$ in units of inverse Hubble times
$t_H^{-1}$, and $t$ is in units of $t_H$.  As DM is gradually loaded
into the DM bins, their physical extents are scaled-up to mimic the
larger core-radius of larger clouds (see \S5.1.1 of Paper 1 for
scaling relations).  Mass is added to the baryonic bins in the same
way, however, the radial positions are not altered directly, but
allowed to evolve according to the pressure gradients and total
gravity of the cloud.  This approximation may result in a slightly
more concentrated distribution of baryons than if the gas was
originally associated with the accreting sub-halos.  But the inward
migration of DM would tend to pull the gas inward anyway.  It appears
that the only significant difference between the model and the envisioned
reality may be a weaker discontinuity in the model between the
outflowing wave from the original ionization of the cloud and the
infalling gas from the periphery.  But the relatively mild difference
between fixed halo, and grown halo evolution suggests that a
thoroughly accurate treatment of this problem would present a final
profile much like the grown halos.

Results for some of the halos are shown in Fig. \ref{fig-g-ng}, where
the solid density profiles represent the ``grown'' halos, and the
accompanying dashed lines represent fixed halos of the same circular
velocity.  In general, the resulting halos have a modestly lower
central density and a modestly greater density outside the core.  In
this way, clouds with central bin velocities up to $v_c = 33.5 $ \kms\
were generated with no central baryon collapse, three more velocity
bins than with fixed halos.  For still larger halos, however, the
innermost bins collapse, indicating the plausible formation of stars.
The bin with center 37.6 \kms\ collapsed at very low redshift,
indicating the possibility of very young void galaxies forming from
such clouds.  Thus, the growth of void clouds could explain the
existence of blue compact galaxies at voids edges
\citep{Pustilnik:95}, with halo circular velocities $v_c \ga 35$
\kms.

\begin{figure}[h!] 
\centering
\epsscale{01.0} 
\plotone{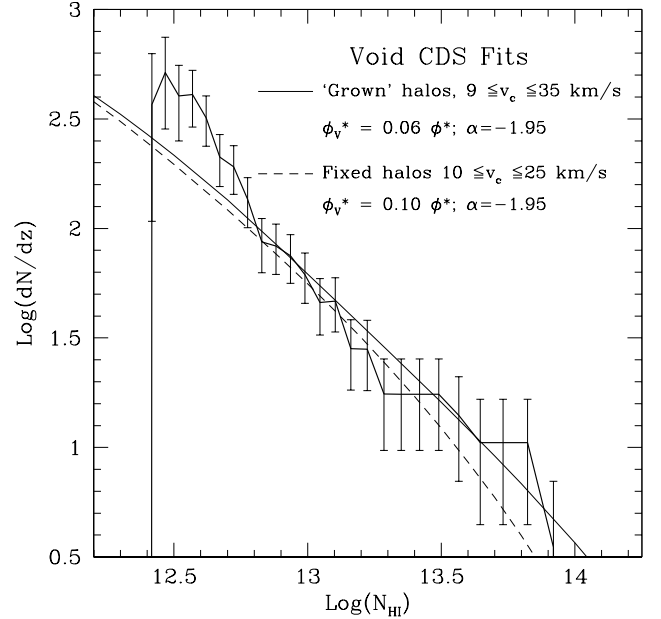}
\caption{\label{fig-CDS-g-ng} Predicted CDS using an HVDF with slope
parameter $\alpha=-1.95$ for clouds evolved with fixed halos (dashed
line), compared with halos which grow according to Eq. \ref{eq-grow}.
The latter halos are detectable in 4 more bins than the former, one
centered at 9.4 km/s, and three at 26.6, 29.9 and 33.5 km/s.  The
normalization for the grown clouds was $\phi_V^* = 0.06~ \phi^*$, and
for fixed halos, $\phi_V^* = 0.1 ~\phi^*$. }
\end{figure}

For a given normalization $\phi_V^*$, extending the halos to larger
circular velocities in this way results in a larger line density at a
given column density.  As a result, the best fit to the observed CDS
requires lowering the normalization to $\phi_V^* \simeq 0.06 ~\phi^*$.
The resulting CDS, using the grown halos with HVDF slope parameter
$\alpha=-1.95$, is presented in Fig. \ref{fig-CDS-g-ng} as the solid
line; the dashed line represents the fixed halos with $\phi_V^* = 0.1
~\phi^*$.  Note that the quality of the fit is greatly improved at
large column densities.  The excess at small column density is still
expected, due to the sub-clustering of small halos in the flanks of
major clouds.  Modeling this component requires semi-analytic work
beyond the scope of the present paper, and is deferred.

If the mass function is integrated over the range of model halo
velocities which are detectable, the mass density in voids is,
$\Omega_V \ge 0.022$, for $\phi_V^* = 0.06 \, \phi^*$ and $R_t = 500 \,
X_v ~\kpc$.  This is 78\% of the mass density for fixed halos.  The
reason for this difference is first, the lower normalization, but this
is partially compensated by the larger halos which emerge from the
grown halo models.

\section{Doppler Parameters of Model Absorbers} \label{sec-bval}

\begin{figure}[h!] 
\centering 
\epsscale{1.0} 
\vspace{-0.5cm}
\plotone{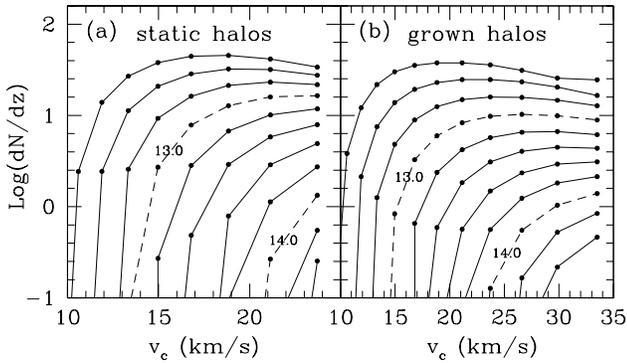}
\vspace{-3.5cm}
\caption{\label{fig-tdnrx} Panels $a$ and $b$ show the cumulative line
density (greater than a given column density) as a function of the
halo velocity bin for fixed, and grown halos, respectively.  The top
line is the minimum detected column density, $N_{HI}=10^{12.4}
$ \cmtw, with subsequent lines representing columns larger by 0.2 in
$\log{N_{HI}}$.  A few of the lines are labeled by the log of their
column density.  Low column density clouds are produced by clouds with
a wide range of halo velocities.}
\end{figure}

Recall that the diffuse absorbers expected from N-body simulations
predict large Doppler parameters, and that this was one of the reasons
that this author abandoned the diffuse cloud model
(\S\ref{sec-intro}).  The evolved baryon distributions pertinent to a
given halo model have a wealth of detail in the program output of the
1-D simulations.  This section investigates the Doppler parameters of
the model absorbers.  Fig. \ref{fig-tdnrx} shows the cumulative line
density $d{\mathcal N}/dz$ contributed by each of the model halos, as
normalized by $0.1 \phi^*$ for the fixed halos (panel $a$), and $0.06
\phi^*$ for grown halos (panel $b$); both with slope parameter
$\alpha=-1.95$.  Note that low column density absorbers are produced
in a wide range of cloud models.

\begin{figure}[h!] 
\centering
\epsscale{1} 
\plotone{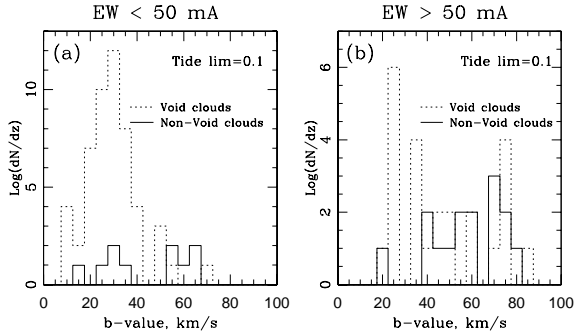}
\vspace{-4cm}
\caption{\label{fig-bval2} Cloud Doppler parameter histograms from
\citet{Penton:00a} GHRS data for void (dotted) and non-void (solid
line) catalogs (using $z \le 0.036$), for clouds with ${\mathcal W} <
50$ m\AA\ (left-hand plot), and ${\mathcal W} > 50$ m\AA\ (right).
For low-EW clouds there is a narrow range of $b$-values, while for
larger clouds, $b$-values appears to vary over a wide range.  Notice
the difference in vertical scales between panels $a$ and $b$.  }
\end{figure}

Figure \ref{fig-bval2} shows histograms of void (dotted) and non-void
(solid lines) clouds for ${\mathcal W} \le 50$ m\AA\ (panel $a$) and
${\mathcal W} > 50$ m\AA\ (panel $b$), for the low-redshift sample, $z
\le 0.036$.  Panel $a$ shows that low EW void clouds have remarkably
uniform $b$-values, with a near-symmetrical profile with a maximum
at $\sim 30$ \kms, while panel $b$ shows that of the higher EW clouds,
non-void clouds predominate only at higher b-values.  That is, almost
half of the clouds with ${\mathcal W} \ge 50$ m\AA\ and $v_c \ge 40$
\kms\ are void clouds.  

Fig. \ref{fig-tdnrx}$b$ shows that, for grown clouds, a cloud of
column density $N_{HI}=10^{12.4} $ \cmtw\ can be produced by clouds of
circular velocity $10.6 \la v_c \la 35$ \kms.  For purposes of
illustration, I now undertake to closely inspect the thermal and
velocity structure of model absorption lines produced by the extremes
of halos producing this column density.

I estimate the core temperature of clouds with halo velocities
$v_c=10.6$ and 33.5 \kms\ using the method of \S\ref{sec-T-b}. The
smaller cloud is found to have a weighted temperature of 7863 K, while
the larger cloud has a temperature of 5375 K.  This can be understood
by reference to Fig. \ref{fig-lagr-hui}, for the larger cloud has a
much larger impact parameter than the smaller cloud ($r_p \simeq 500
~\kpc$ \vs $\sim 55$ \kpc, respectively), and hence has lower average
densities.

There is apt to be large-scale turbulence in the clouds following
reionization, especially toward the center where the density, and its
gradient remain high, and the potential for convective instability is
large.  A plausible upper-limit for a cloud's turbulent velocity is
the cloud virial velocity.  The virial temperature is,
\begin{equation}
T_{vir} =  \frac{1}{2} \frac{v_c^2 \mu m_H}{k_b},
\end{equation}
and may be converted to a virial $b$-value by $b_{vir}=\sqrt{0.0165
T_{vir}}$.  The virial temperatures of the two clouds ($T_{vir} =
6776$ K, and 67,756 K, for the smaller, and larger clouds,
respectively) may contribute to the Doppler parameter.  For the small
10.6 \kms\ cloud, the addition to $b$ contributed by turbulence is apt
to be small since the $T_{vir}$ is less than the thermal temperature.
For the large cloud, the turbulent component could dominate, depending
on the impact parameter (hence column density).  Since the impact
parameter of the large $N_{HI}=10^{12.4}$ \cmtw\ cloud is large, the
sightline won't penetrate the virializing portion of the cloud ($r_p
\la 125$ kpc).  Instead, the velocity field of the expanding cloud may
be an important addition.  The LOS velocity difference within the FWHM
of the line is found to be 25 \kms\ over the LOS distance,
$l_{1/2}=534 ~\kpc$, yielding $b_{vel}\simeq 15 $ \kms.  If this
material were in the Hubble flow, the FWHM velocity would have been
$\sim 40 $ \kms, showing the physical effect of the negative peculiar
velocities previously noted for large clouds (see
Fig. \ref{fig-lagr-vel4}).

The $b$-value associated with the smaller $N_{HI}=10^{12.4}$ \cmtw\
cloud is then $b = \sqrt{b_{therm}^2 + b_{turb}^2} \le 15.5 $ \kms,
where the $b_{vel}$ component is ignored since it is found to be less
than $2 $ \kms\ within the FWHM of the line.  The direction of the
inequality in the equation stems from the fact that $v_{turb}$ is an
upper limit for bulk motions in these absorbers.  The lower limit is
the thermal $b$-value; $b= 11.4$ \kms.  For the larger cloud, $b =
\sqrt{b_{therm}^2 + b_{vel}^2} \ge 17.7 \kms$, where the turbulent
(virial) motions of the cloud have not been included since only the
outer cloud is probed.  This velocity is a lower limit because some
turbulent motions are likely.  

Larger clouds, such as $N_{HI}=10^{13.5}$ \cmtw\ (${\mathcal W} \simeq
130$ m\AA), have averaged temperatures on order $\sim 16,900$ K
($b_{therm}=16.7$ \kms), produced by halos of velocity $17 \la v_c \la
33$ \kms.  They give total $b$-values $b\la 24$ to $b \la 37$ \kms,
where I have ignored the cloud expansion broadening, which is dwarfed
by possible turbulent motions.  The impact parameters for these
systems, $120 \la r_p \la 180 ~\h_{75}^{-1}$ \kpc, may probe virializing
regions.

These values agree well with the range of $b$-values presented in
\citep{Dave:01} (see their Fig. 6$a$).

In comparing the analysis of $b$-values in \citet{Dave:01} to those in
this paper, it is interesting to note that though the former predict
higher temperatures for a given gas density (see
Fig. \ref{fig-lagr-hui}), the thermal $b$-values as a function of a
given column density (Eq. \ref{eq-temp}) are smaller.  Table 1 shows a
compilation of thermal $b$-values; column 1 is the log column density;
the next column shows the \citet{Dave:01} thermal $b$-values; third
are those of the present study.  It would appear that at the same
column density, the former must have substantially lower \hone\
densities (by a factor of $\sim 3$ to 5) than the discrete clouds
considered here, and therefore have proportionately larger integration
paths.  The irony should not be missed; the absorbers of
\citet{Dave:01} are supposed to occupy a dense environment, but are
less dense than the void absorbers modeled in the present effort.

\small
\begin{table}[h!] 
\begin{center}
\caption{Predicted Thermal Doppler Parameters}
\medskip
\begin{tabular}{ccc}
\tableline
\tableline
$\log N_{HI}$ \,(\cmtw) & $b_{therm}$ (\kms)  & $b_{therm}$ (\kms)  \\
   &   \citet{Dave:01} & Manning \\
\tableline
12.5  &  9.2  &  11.3  \\
13.0  & 11.7  &  13.8  \\
13.5  & 14.9  &  16.7  \\
14.0  & 19.0  &  21.1  \\
\tableline
\tableline
\end{tabular}
\end{center}
\medskip
\end{table}
\normalsize

I have shown that centrally condensed clouds based on sub-galactic
halos produce Doppler parameter distributions consistent with observed
clouds \citep{Dave:01}.  Perhaps, however, good agreement is
\emph{not} expected since \citet{Dave:01} purport to study
predominantly non-void clouds, while the present study models void
clouds.  On the other hand, if the lines of sight covered in their
paper are like those in \citet{Penton:00a}, then Fig. \ref{fig-tdnrx}
would indicate that most clouds, especially those of lower column
density, are in fact void clouds, and substantial agreement with the
echelle observations would be expected.

However, it is unlikely that one could decide between giant sheets and
discrete, sub-galactic absorbers on the basis of Doppler parameters
alone.  The conclusion to be taken away from this section is that
discrete halos produce results that are consistent with observations
of resolved Doppler parameters in low-redshift \lya\ clouds.

\section{Summary} \label{sec-summary}

In the introduction, two possibly problematic aspects of recent
N-body/hydro simulations were noted; first, the fact that \lya\
absorbers are expected to be produced in diffuse clouds by a
``fluctuating Gunn Peterson'' effect \citep{Croft:98, Dave:01}, rather
than by absorption in discrete halos.  Second, that low-redshift absorbers
are expected in diffuse filamentary structures surrounding galaxy
concentrations, but not in voids.

The analysis in \S\ref{sec-Dave} showed that indeed, diffuse
structures could not provide an explanation for the void absorbers.
The analysis in the balance of the paper showed that discrete halos
may restrain the baryons from evaporating to a degree sufficient for
absorption systems to be produced that are consistent with
observations.  This implies that it is \emph{not} a good approximation
to assume that small halos would not retain a significant quantity of
their baryons following reionization.

However, the problem goes farther than just contesting the N-body
\lya\ cloud simulations.  For the current halo paradigm, NFW, also
fails to explain the absorbers since those halos small enough to
escape star formation are not massive enough to hold baryons against
photo-evaporation.  In short, a more massive halo is needed.

I have shown that an isothermal halo \emph{can} explain the absorbers,
within the cosmological parameters investigated here.  Some might
argue that the ISOT halo is not physically realistic since they cannot
be virialized.  However, I am not proposing that ``ISOT halos'' are
totally virialized.  Instead, I propose (as noted in
\S\ref{sec-nature-init}) that the isothermal density profile is a fair
description only of the total mass distribution.

The NFW halo is said to be universal, but when applied to rich
galaxy clusters, it fails to explain the mass distribution; analysis
of galaxy-galaxy correlations by \citet{Seldner:77} showed that the
distribution of galaxies ($n_{gal} \propto R^{-2.4}$) around rich
clusters extends to $15 ~\h^{-1}$ Mpc, without a feature that would
suggest the expected sharp decline in density beyond $R_{max}$ ($\sim
0.8$ Mpc), or at the virial radius ($\sim 2$ Mpc).  Recent results of
\citet{McKay:02} substantiate this claim using isolated groups of
galaxies.  They find $n_{gal} \propto r^{-2.1}$, and a LOS velocity
dispersion $\sigma_r$ that shows no significant variation between
apertures of 133 to $670 ~\h_{75}^{-1} ~\kpc$ (whereas $r_{vir}
\approx 250 ~\h_{75}^{-1} ~\kpc$).  These results appear to support an
isothermal mass distribution -- either truncated at a very large
radius, or untruncated.

Evolved ISOT baryon profiles whose halos were assumed to be
un-evolving, were found to be more compact, and tended to form stars
at velocities $v_c \ga 25$ \kms, leaving a gap between the largest
model clouds and smallest galaxies and CHVCs.  ``Grown'' halos,
however, allowed clouds with circular velocities up to $v_c \simeq 35$
\kms\ without forming stars, virtually closing the gap between model
clouds and observed galaxies.  The profiles which result from a
gradually growing halo produce a significantly better match to the
shape of the CDS, though the excess at low EW does, and should remain
because the sub-halos are still there affecting the cloud cross
section to absorption at low column density (though their effects
cannot be modeled with a 1-D simulation).

I make the following conclusions:
\begin{enumerate}

\item N-body/hydro simulations fail to predict the observed
void clouds; diffuse sheets cannot explain the absorbers.
 
\item The NFW halo must be rejected as an absorber model on a similar
basis; it cannot explain the observed void clouds if ``NFW'' is taken
to adequately represent the mass distribution \emph{beyond} the virial
radius.

\item A halo more massive than NFW is needed to explain observations.
I find that a model consistent with a non-singular inverse square DM
density profile, with a large truncation radius, is consistent with
the data.

\end{enumerate}
\section{Acknowledgments}
I thank my advisor Hyron Spinrad for his encouragement and support
during my stay at Berkeley.  I am grateful to Christopher McKee for
his incisive inspection of my work and for many useful suggestions.  I
wish to express my appreciation to the anonymous referee for helpful
comments.  My research was funded in part by NSF grant AST-0097163,
and the Department of Astronomy at the University of California,
Berkeley.

\end{document}